\documentclass[twocolumn]{aastex631}

\shorttitle{LMC Bar Geometry}
\shortauthors{Rathore et al.}

\usepackage{multirow}
\usepackage{csquotes}

\begin{document}
\title{Precise Measurements of the LMC Bar's Geometry With Gaia DR3 and a Novel Solution to Crowding Induced Incompleteness in Star Counting}

\correspondingauthor{Himansh Rathore}
\email{himansh@arizona.edu}

\author[0009-0009-0158-585X]{Himansh Rathore}
\affiliation{Department of Astronomy and Steward Observatory, University of Arizona, 933 North Cherry Avenue, Tucson, AZ 85721, USA}

\author[0000-0003-1680-1884]{Yumi Choi}
\affiliation{NSF National Optical-Infrared Astronomy Research Laboratory, 950 North Cherry Avenue, Tucson, AZ 85719, USA}

\author[0000-0002-7134-8296]{Knut A.G. Olsen}
\affiliation{NSF National Optical-Infrared Astronomy Research Laboratory, 950 North Cherry Avenue, Tucson, AZ 85719, USA}

\author[0000-0003-0715-2173]{Gurtina Besla}
\affiliation{Department of Astronomy and Steward Observatory, University of Arizona, 933 North Cherry Avenue, Tucson, AZ 85721, USA}

\begin{abstract}
We present new measurements of the two-dimensional (2-D) geometry of the LMC's stellar bar with precise astrometric observations of red clump stars in Gaia DR3. We develop a novel solution to tackle crowding induced incompleteness in Gaia datasets with the Gaia BP-RP color excess. Utilizing the color excess information, we derive a 2-D completeness map of the LMC's disk. We find that incompleteness biases the bar measurements and induces large uncertainties. With the completeness-corrected 2-D red clump map, we precisely measure the LMC bar's properties using Fourier decomposition. The bar radius (semi-major axis) is $R_{bar} = 2.13^{+0.03}_{-0.04}$ kpc, and its position angle is $121.26^{\circ} \pm 0.21^{\circ}$. The bar's strength as quantified by the Fourier bi-symmetric amplitude is $S_{bar} = 0.27$, indicating that the LMC has a significant bar perturbation. We find the bar has an axis ratio of $0.54 \pm 0.03$, and is offset with respect to the center of the outer disk isophote at R $\approx$ 5 kpc by $0.76 \pm 0.01$ kpc. These LMC bar properties agree with a hydrodynamic model where the SMC has undergone a recent direct collision with the LMC. We compare the LMC's bar properties with other barred galaxies in the local universe, and discover that the LMC is similar to other barred galaxies in terms of bar-galaxy scaling relations. We discuss how our completeness correction framework can be applied to other systems in the Local Group.
\end{abstract}

\keywords{\href{http://astrothesaurus.org/uat/903}{LMC (903)}; \href{http://astrothesaurus.org/uat/2364}{Galaxy bars (2364)}; \href{http://astrothesaurus.org/uat/1043}{Astronomical methods (1043)};  \href{http://astrothesaurus.org/uat/767}{Hydrodynamical simulations (767)}; \href{http://astrothesaurus.org/uat/416}{Dwarf galaxies (416)}}

\section{Introduction} \label{sec:intro}

The stellar bar of the LMC is one of the most peculiar features of this massive Milky Way satellite galaxy. Unlike bars in most gas rich galaxies of the local volume, the LMC's bar is offset from the isophotal center of the disk \citep{deVFreeman72, vdM2001} and is absent in the LMC's interstellar medium (ISM) \citep{Stavely-Smith2003}. Moreover, the bar is significantly more dominant in the older stellar populations (age $>$ 1 Gyr) as compared to the younger stellar populations (age $<$ 100 Myr) \citep{Harris2009, ElYoussoufi2019, Mazzi2021, Luri2021}. The nature of the LMC's bar is a topic of debate. Precise characterization of the geometry and strength of the LMC's bar is needed to place the LMC in context with other barred galaxies in the local universe and understand the role of bars in general in the evolution of low mass galaxies. 

Early work questioned whether the LMC's bar is comparable to standard galactic bars and actually embedded within the LMC disk. Using observations from micro-lensing surveys \citep{Gould1995}, \cite{ZhaoEvans2000} suggested that the bar is an un-virialized structure that is above the plane of the LMC disk. Using UBVI photometry from the Magellanic Clouds Photometric Survey \citep{Zaritsky2004b}, \cite{Zaritsky2004} suggested that the bar is a manifestation of viewing a tri-axial stellar bulge. Furthermore, based on Cepheid distances, \cite{Nikolaev2004} suggested that the LMC's bar is levitating with respect to the disk plane by as much as $0.5$ kpc. 

As distance measurements have become more accurate, a different picture has emerged. Using distances to red clump stars, it has been found that the bar actually resides in the disk plane \citep{Subramaniam2009}. Using state-of-the-art observations from the Survey of MAgellanic Stellar History \citep[SMASH][]{Nidever2017}, \cite{Choi2018a} found that the bar is consistent with being in the disk plane, but is likely tilted with respect to the disk by as much as $\sim 15^\circ$. \cite{Haschke2012} also infer that the LMC's bar resides in a different plane relative to the disk. The observation that the bar is tilted helps to understand how earlier studies could conclude that the LMC's bar was not in the disk plane. 

It has been proposed that the LMC bar tilt is caused by a recent direct collision between the SMC and the LMC, where the LMC originally possessed a centered, in-plane stellar bar that was distorted during the collision \citep{Besla2012}. If this is true, then the LMC bar must currently have 2-D geometric properties (like radius, axis ratio) and a bar strength similar to that expected for a galaxy of this mass scale.   

In addition to the nature of the LMC's stellar bar, the connection between the bar and surrounding ISM is not well understood. HI density and velocity maps of the LMC indicate no signature of a bar \citep{Stavely-Smith2003}, which is in contrast to what is observed for most massive gas-rich barred galaxies where the bar is evident both in HI and other ISM tracers \citep[e.g.][]{Bosma1978, Fathi2005, Lopez-Coba2022}, even when the bar is offset from the disk center \citep{Athanassoula1989}. The absence of the LMC's bar in its ISM gives credence to the idea that the LMC bar is not a standard dynamical bar, making the need to accurately characterize the LMC bar's properties even more important.    

Various attempts have been made to measure the 2-D geometry of the LMC's stellar bar, which is quantified by the bar position angle, radius (semi-major axis), ellipticity (axis ratio), and offset of the bar center with respect to the center of the stellar disk isophotes. Table \ref{tab:lit_bar} provides a survey of estimates of these properties of the LMC's bar in literature. The table illustrates that there is little consistency among different works. For example, the radius of the bar differs between studies by as much as a factor of two. Such a high uncertainty is also an impediment towards understanding the origin of the peculiar properties of the bar. For example, an accuracy of at least 0.5 kpc is needed for parameters like the bar radius to accurately determine the torques on the bar from external perturbers like the SMC (H. Rathore et al. 2024(b), in prep).

Precise measurement of the LMC bar's geometry has been mainly hindered by the following challenges:

\begin{table*}
    \centering
    \caption{Literature survey of the estimates of the LMC bar's geometric parameters}
    \begin{tabular}{c c c}
    \hline
    \hline
     Bar Parameter & Value & Reference  \\
     \hline
     \multirow{4}{6em}{Radius [kpc]} & 1.5 & \cite{deVFreeman72}$^{a}$ \\
     & 1.5 & \cite{ZhaoEvans2000}\\
     & 3.0 & \cite{Choi2018b}$^{a}$\\
     & 2.3 & \cite{Arranz2024}\\
     \hline
     \multirow{4}{6em}{Axis ratio} & 0.27 & \cite{deVFreeman72}$^{b}$\\
     & 0.33 & \cite{ZhaoEvans2000}$^{b}$\\
     & 0.33 - 0.57 & \cite{vdM2001}\\
     & 0.46 & \cite{Choi2018b}\\
     \hline
     \multirow{3}{6em}{Position Angle [$^\circ$]} & 120 & \cite{ZhaoEvans2000}\\
     &112 - 126 & \cite{vdM2001}\\
     &154.18 & \cite{Choi2018b}\\
     \hline
     \multirow{2}{6em}{Offset$^{c}$ [kpc]} & 0.4 & \cite{vdM2001}\\
     & 0.87 & \cite{ZhaoEvans2000}\\
     \hline
    \end{tabular}
    \tablenotetext{}{\textbf{Note:} This table is for illustration purposes only, and is by no means meant to be exhaustive. There is a lack of consistency in the determination of bar parameters across different works. $^{(a)}$ Give their measurements in degrees, which we convert to length units by using a distance to the LMC of 49.9 kpc \citep{deGrijs2014}. $^{(b)}$ Give the dimensions of the bar, which we use to determine the axis ratio. We were not able to find meaningful error estimates of the listed quantities in our literature survey. $^{(c)}$ Bar offset is the difference between the center of the bar and the center of the outer disk (R $\approx$ 5 kpc) isophote.}
    \label{tab:lit_bar}
\end{table*}

\begin{itemize}
    \item \textbf{Incomplete stellar photometry due to crowding}: the stellar field in the central region of the LMC is highly crowded, which significantly affects measurements of the bar structure that are based on resolved stars \citep[e.g.,][]{Choi2018a}. Crowding manifests as less complete stellar photometry in the central regions of the LMC, which leads to a systematic underestimate of the stellar density field in the bar region. Incomplete stellar photometry prevents an accurate measurement of the dynamical strength of the bar, which particularly hinders the comparison of the LMC to other barred galaxies. To the best of our knowledge, the LMC bar's strength has not been measured before with the LMC's stellar density distribution.
    \item \textbf{Lack of an accurate geometric framework to measure the bar parameters}: majority of the methods used in literature have used either iso-ellipse fitting on star-count maps to characterize the bar \citep[e.g.][]{vdM2001}, or fitting of two component (bar $+$ disk) models \citep[e.g.][]{ZhaoEvans2000, Choi2018b}. However, separating the bar from the rest of the disk and defining the end of the bar is a significant challenge in these methods. In addition, some bar parameters, like the bar radius, are sensitive to the choice of the prior adopted \citep[e.g.][]{Choi2018b}. These issues make it hard to compare measurements across different studies, or with numerical simulations. A geometric framework is needed that yields accurate measurements of multiple bar parameters without relying on prior assumptions of bar/disk mass profile, and can be extended to different types of stellar tracers as well as compared with numerical simulations.
\end{itemize} 

In this work, we utilize observations of red clump stars in Gaia DR3 to quantify the 2-D structure of the LMC's bar, defined by the position angle, radius, dynamical strength, axis ratio and offset. The offset is defined as the separation between the bar center and the center of the outer disk isophote at R $\approx$ 5 kpc. A standard way of accounting for crowding induced incompleteness is through artificial star tests (ASTs) \citep[e.g.][]{Olsen2003}. However, given the large on sky extent of the LMC, the ASTs are computationally expensive and beyond the scope of the present work. Moreover, the raw Gaia fields are not publicly available at this point, which are needed to perform the ASTs. We present a novel, computationally in-expensive way to account for the crowding-induced incompleteness in the LMC's central regions by utilizing the Gaia BP-RP color excess as a tracer of the crowding level. We present a Fourier decomposition framework, motivated by N-body simulations \citep{AM2002}, to measure the bar parameters in a purely geometric way, without reliance on model fitting. Determination of 3-D geometric parameters of the LMC's bar, like its tilt, is beyond the scope of this work.

Finally, we shall compare our measured LMC bar properties with those of numerical models in which the SMC has undergone a recent direct collision with the LMC \citep{Besla2012} in order to assess the validity of such a hypothesis.

We structure our manuscript as follows. In section \ref{sec:gaiadata}, we 
present the Gaia DR3 observations of red clump stars and motivate a novel solution to the crowding-induced incompleteness issue. In section \ref{sec:comp_map}, we derive a completeness map of the LMC, i.e. completeness as a function of location in the LMC's disk. In section \ref{sec:fourier}, we present the framework of Fourier decomposition for determining the LMC's bar properties. In section \ref{sec:lmc_bar}, we present our measurements of the 2-D geometric parameters of the LMC's bar. In section \ref{sec:discussion}, we compare our measurements to numerical simulations and place the LMC in context with other barred spiral galaxies. We summarize our work in section \ref{sec:conclusion}. Throughout this manuscript, we assume a distance to the LMC of $D_{LMC} = 49.9$ kpc \citep{vdM2001, deGrijs2014}.

\section{Gaia DR3 Data for the LMC and a Novel Solution to Incompleteness} \label{sec:gaiadata}

Gaia has ushered a new era of galaxy dynamics by providing precise astrometry for stars in the Milky Way and nearby galaxies like the LMC (e.g. \cite{Luri2021} and references therein). These data bring novel opportunities for precise characterization of the structure and kinematics of the LMC, including its stellar bar \citep[e.g.][]{Zivick2019, Choi2022, Dhanush2024, Arranz2024}. 

We utilize red clump stars in Gaia DR3 \citep{Gaia2016, Lindegren2021, Gaia2023} for precise geometric characterization of the LMC's bar. The LMC's bar predominantly consists of older stellar populations \citep{Harris2009, Mazzi2021, Luri2021}, thus red clump stars are representative of the bar. Moreover, red clump stars occupy a narrow and a well defined region in the Hertzsprung-Russel diagram \citep{Castellani2000}, which makes them easy to select in Gaia Color-Magnitude diagrams \citep[e.g.][]{Choi2022}. Using red clump stars will also allow us to place our results in context with several other works that have utilized this stellar population to study the LMC's bar and disk  \citep[e.g.][]{Subramaniam2009,Choi2018a,Choi2022}. Red clump stars have a similar mass \citep{Castellani2000}, which makes comparisons with numerical simulations where star particles have equal masses easier. Further, red clump stars are standard candles for distance estimation \citep{Girardi2016}, offering potential for extending our framework to determine the 3-D geometry of the LMC's bar in future analyses.

We follow the selection criteria of \cite{Choi2022} (hereafter C22) for selecting red clump stars in the LMC using the Gaia query interface \footnote{\url{https://gea.esac.esa.int/archive/}}. We describe the exact Gaia query and selection procedure for the red clump stars in Appendix A. At this stage, we have a sample of 1,051,494 red clump stars. We refer to this sample as our parent sample. We utilize the neural network based catalog of \cite{Arranz2023} to evaluate the contamination of Milky Way foreground stars in our sample of LMC stars. We estimate this contamination to be less than 1\% in the LMC bar region (R $< 3^\circ$) for their NN complete sample.

\subsection{Coordinate Systems} \label{sec:coords}

In this work, we utilize two coordinate systems to describe the LMC's disk. These follow from \cite{vdMCioni2001, Salem2015, Mackey2016,Choi2018b}. 

For visualizing the LMC's disk on-sky, we use the LMC tangent plane coordinate system ($\eta$ [$^\circ$], $\xi$ [$^\circ$]). This coordinate system is obtained by projecting the LMC's disk plane on the tangent plane to the celestial sphere at the LMC's kinematic center (known as the gnomonic projection).

For analyzing the geometry of the LMC's bar and for comparing it with numerical simulations, we use the LMC in-plane cartesian coordinate system (x [kpc], y [kpc]). This coordinate system is obtained by defining an infinitely thin plane across the LMC's disk that is inclined to the sky by the LMC's observed inclination about its line of nodes and oriented at the observed position angle of the line of nodes.  

We describe how to obtain these coordinates from the RA - DEC coordinates in Appendix B. 

\subsection{Motivating an Incompleteness Correction Procedure Using the Gaia Color Excess}

The Gaia G filter is almost exactly partitioned into the BP (blue photometer) and RP (red photometer) filters. Ideally, the fluxes in BP and RP filters should approximately add up to the flux in the G filter. In reality, there can be a difference between the total flux in BP $+$ RP filters compared to the flux in G. This difference is quantified by the Gaia color excess $C$, which is the ratio of the total BP $+$ RP flux over the G flux. The ratio is expected to be slightly larger than unity even in ideal observing conditions since the BP and RP filters taken together cover a larger wavelength range compared to the G filter. 

\cite{Evans2018} concluded that stars with a value of $C$ significantly different from unity likely have biased photometry, and they recommend removing such stars. However, \cite{Riello2021} (hereafter R21) found that $C$ has a significant dependence on the BP-RP color, leading to results that are difficult to interpret when it is used as a selection criteria. They derived a corrected color excess $C^\ast$, in which the global color dependence of $C$ is subtracted out. R21 recommend using those stars whose $C^\ast$ value is close to 0.

\begin{figure*}
     \centering
     \includegraphics[height = 0.25\textwidth, width = 0.30\textwidth]{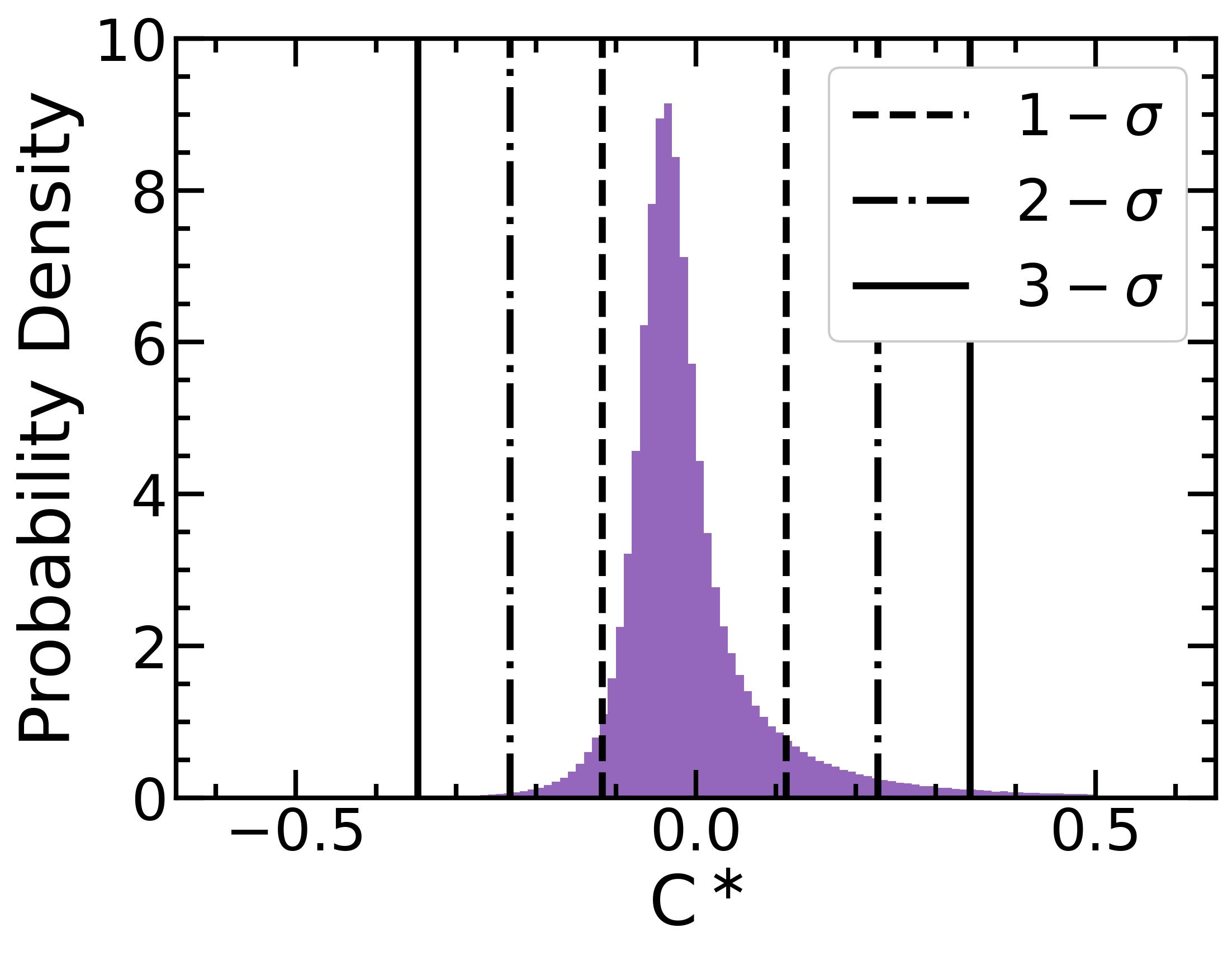}
     \includegraphics[height = 0.25\textwidth, width = 0.34\textwidth]{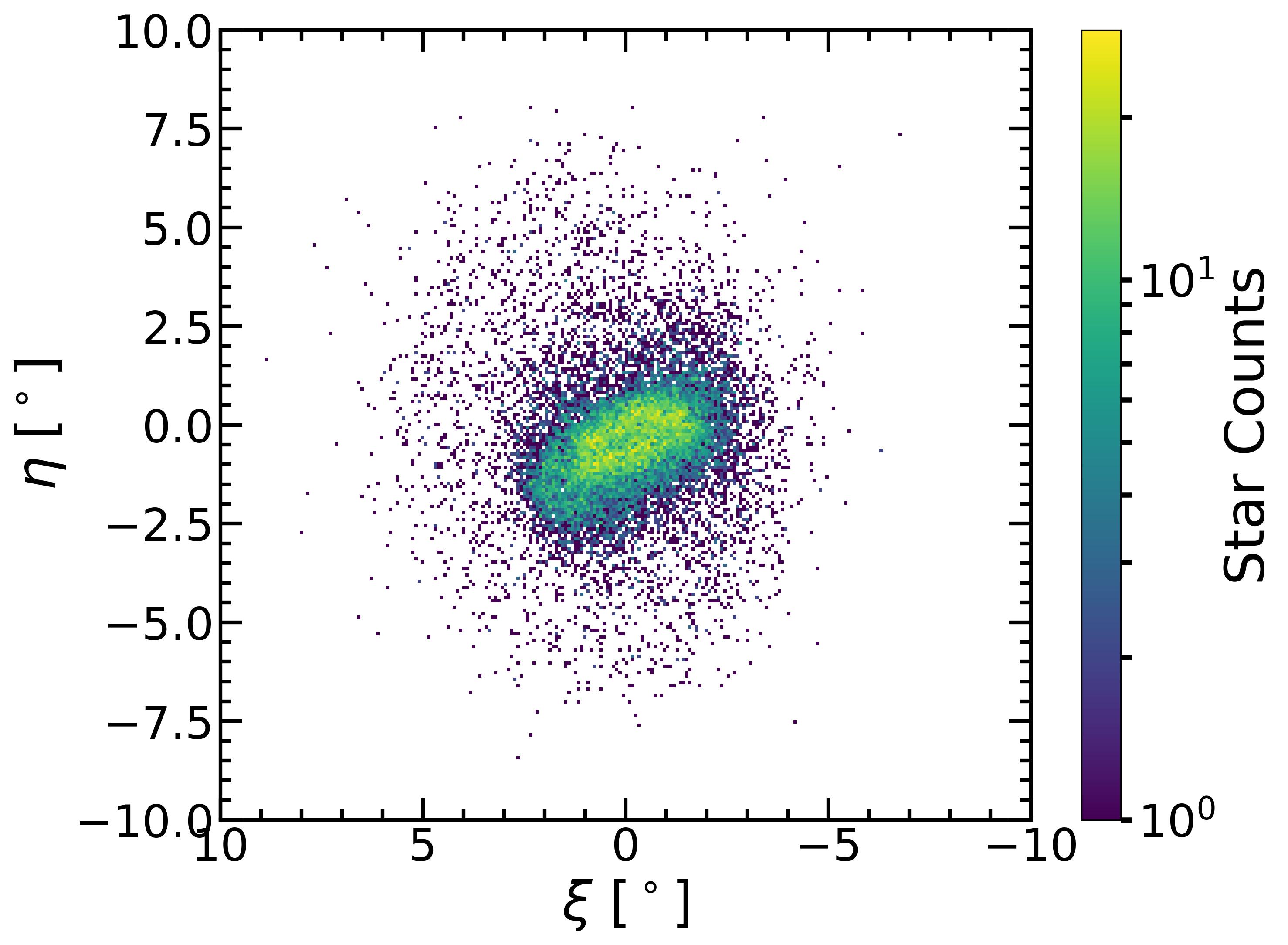}
     \includegraphics[height = 0.25\textwidth, width = 0.34\textwidth]{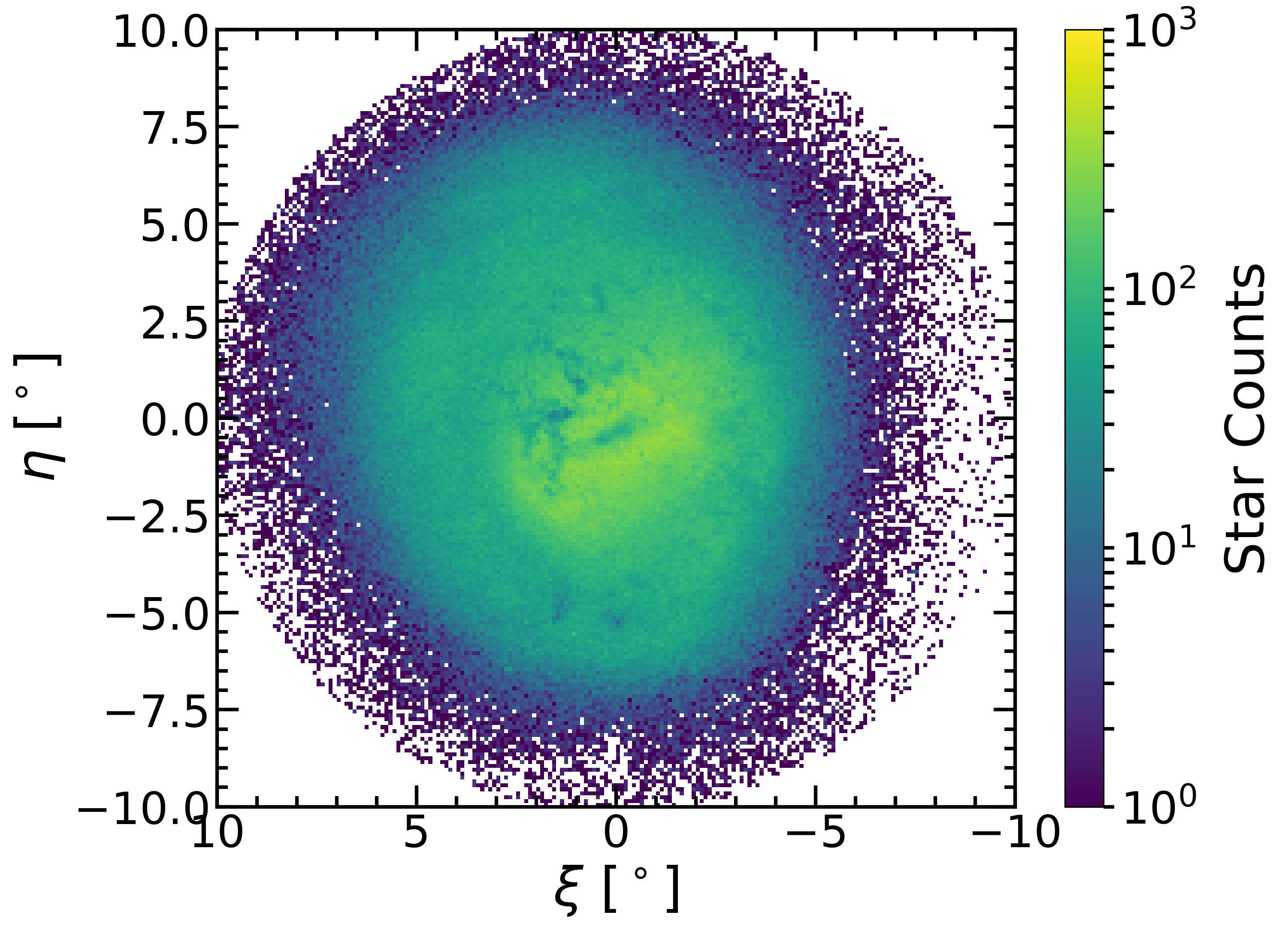}
     \caption{{\em Left Panel}: The probability density distribution of the corrected Gaia BP-RP color excess ($C^\ast$) for the sample of LMC red clump stars obtained by applying the selection criteria of C22 as shown in Appendix A. The region enclosed by the dashed line, dash-dot line and the solid line represent one, two and three standard deviations from the mean $C^\ast$ respectively. {\em Middle Panel}: The spatial distribution of stars residing beyond the $3-\sigma$ distribution as shown in the {\em left panel}, which we remove from our sample. These excluded stars constitute around $1\%$ of the initial count and trace the crowded bar region of the LMC's disk. This indicates that $C^\ast$ can be used as a representative of crowding induced incompleteness in the LMC. {\em Right Panel}: The spatial distribution of the sample of red clump stars that we obtain after applying the selection criteria based on $C^\ast$. The star counts are significantly underestimated in the central region, which is an effect of crowding. We refer to this as the \enquote{Incomplete sample}.}
    \label{fig:cuts}
\end{figure*}

Thus, following C22, we apply the following restriction to our parent sample:
\begin{equation}\label{eq:Cstar_cut}
    |C^\ast| < 3 \sigma_{C^\ast}
\end{equation}
To apply this cut, we construct the distribution of $C^\ast$ (shown in Figure \ref{fig:cuts}, left panel) and remove stars that reside beyond three standard deviations relative to the mean. The spatial distribution of the excluded stars is shown in Figure \ref{fig:cuts}, middle panel. After this cut, we have a sample of 1,041,822 red clump stars, which is smaller than the parent sample by $\sim$1\%.  We refer to this as the \enquote{Incomplete sample}, since star counts in the LMC's central region are underestimated due to crowding (see Figure \ref{fig:cuts}, right panel).  

We observe from the spatial distribution of the excluded stars (Figure \ref{fig:cuts}, middle panel) that they primarily trace the crowded bar region of the LMC, indicating that $C^\ast$ is a good probe of crowding level in the LMC's disk. Below, we explain how $C^\ast$ can be used to inform crowding-induced incompleteness of our selected red clump stars, based on \cite{Voggel2020}, \cite{Hughes2021} and R21. 

The flux measurement process for the G filter is different from the BP and RP filters. The flux in the G filter is obtained using profile fitting to an image with a resolution of $0.4$ arcsec, whereas the fluxes in BP and RP filters are obtained with an effective aperture of size $3.5 \times 2.1$ arcsec$^2$ \citep{Evans2018}. In crowded fields, fluxes within the BP and RP apertures can be contaminated by nearby sources. This contamination will in general be stochastic, and different for each star in the crowded field. Thus, the scatter in $C^\ast$ at a given location in the LMC's disk probes the extent of crowding in that location.

We interpret the standard deviation of the distribution of $C^\ast$ ($\sigma_{C^\ast}$) for stars at a given location in the LMC's disk as an error in color due to crowding. \cite{Olsen2003} (hereafter O03) devised a relation between the error in color due to crowding and the resulting completeness fraction of stars. This enables us to derive a completeness map of red clump stars in the LMC from $\sigma_{C^\ast}$.

\section{Deriving a Completeness Map for the LMC} \label{sec:comp_map}

\begin{figure*}
     \centering
     \includegraphics[height = 0.25\textwidth, width=0.34\textwidth]{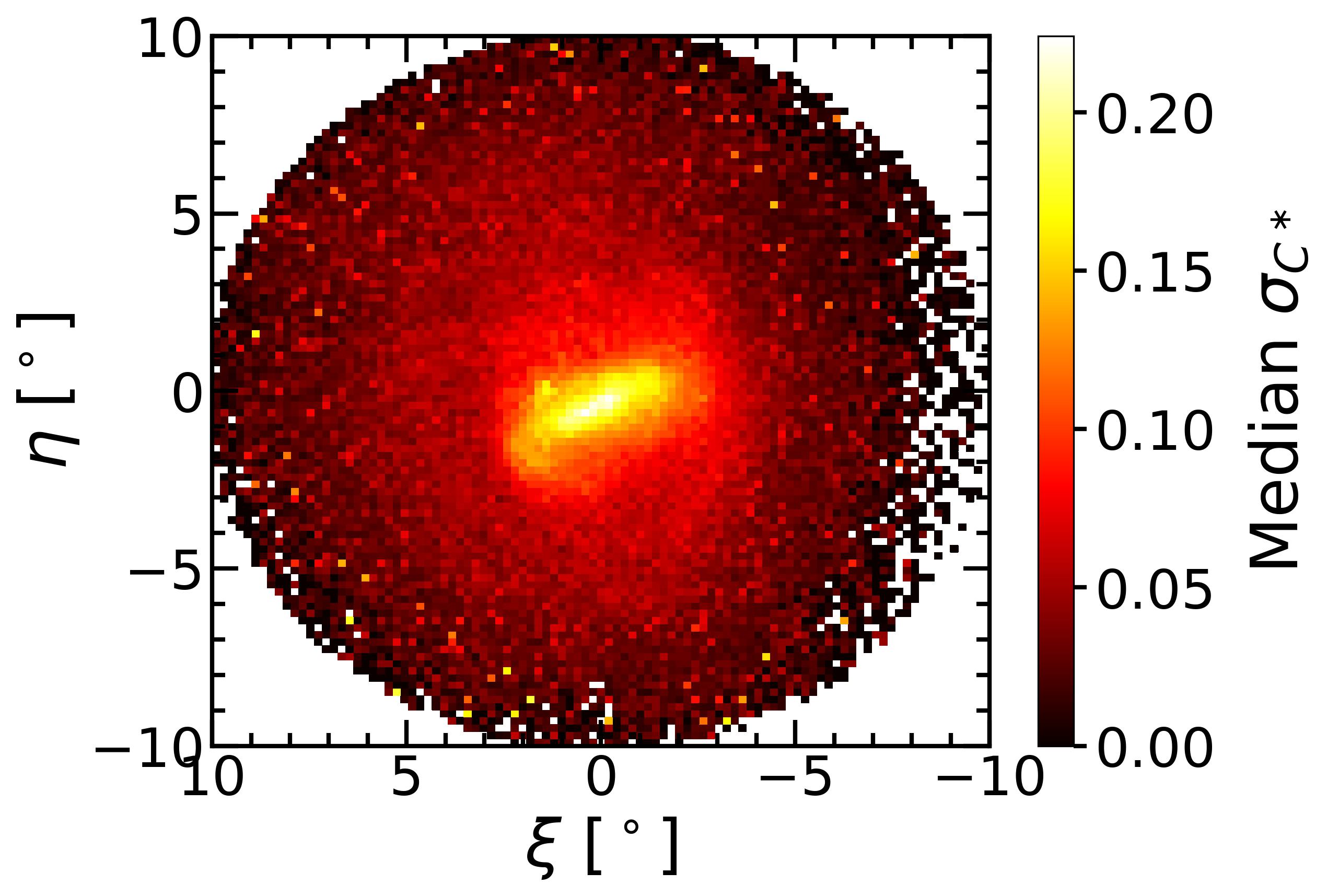}
     \includegraphics[height = 0.25\textwidth, width=0.30\textwidth]{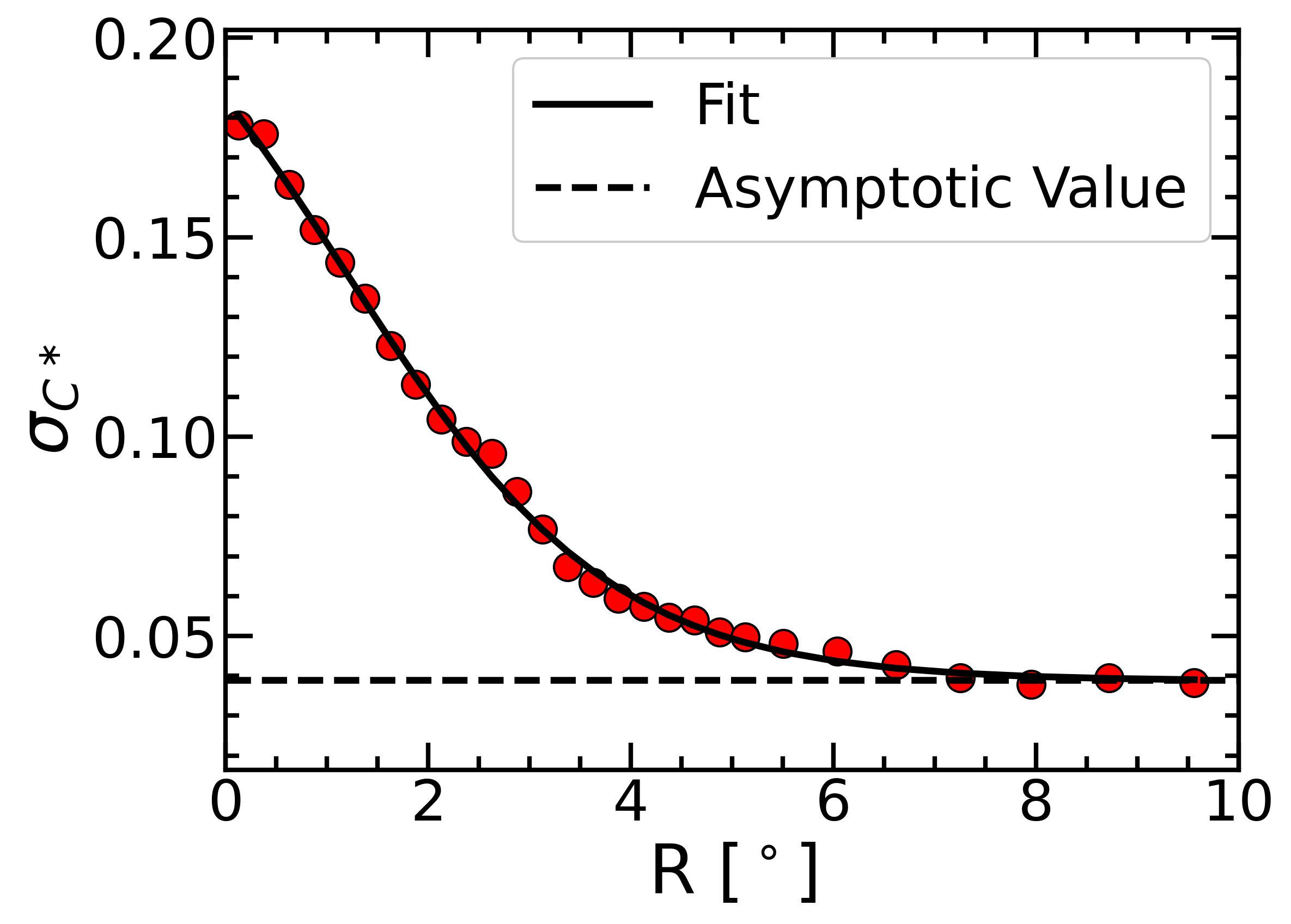}
     \includegraphics[height = 0.25\textwidth, width=0.34\textwidth]{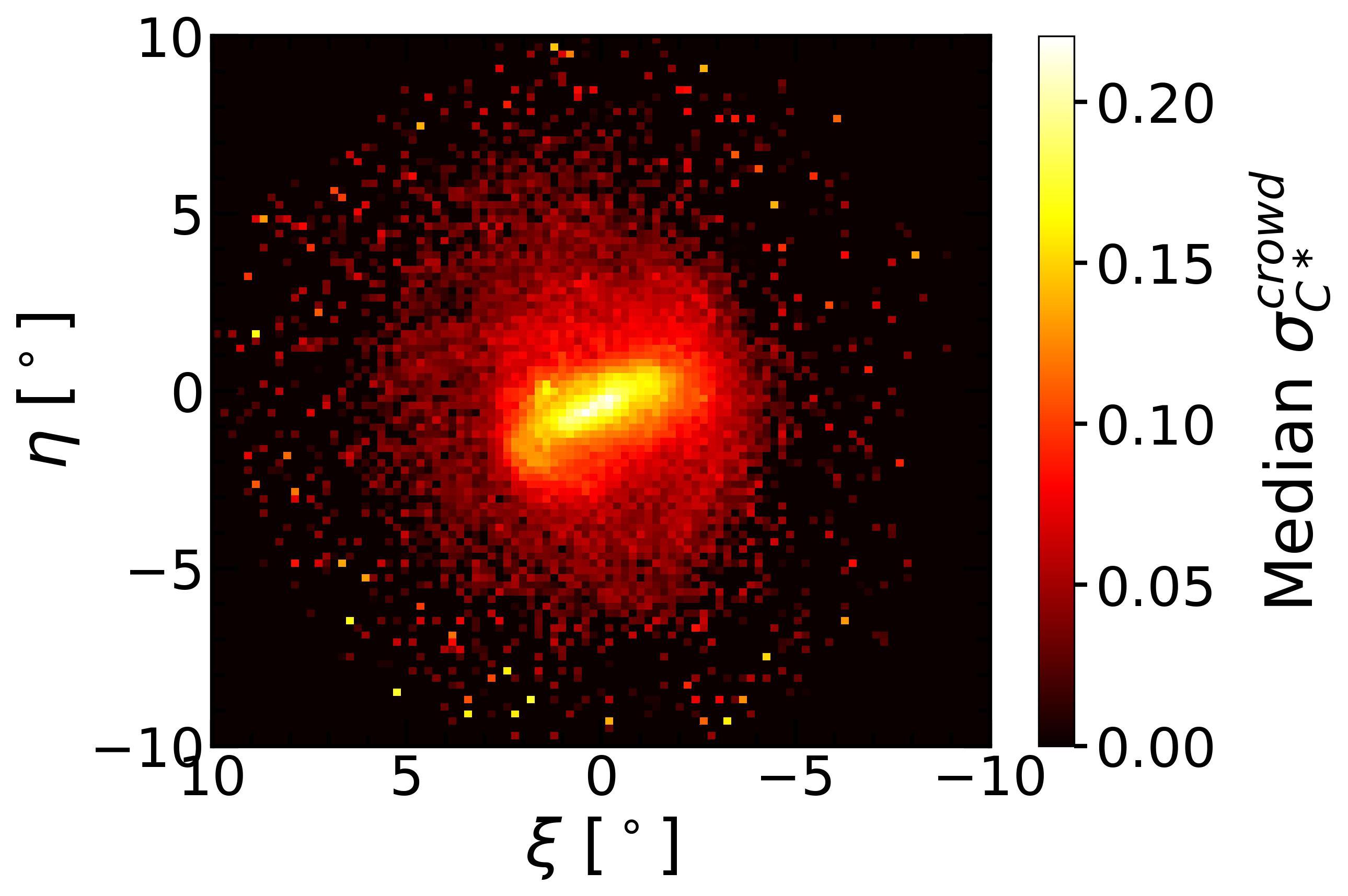}
     \caption{{\em Left Panel}: The spatial distribution of the spread in the corrected Gaia color excess ($\sigma_{C^\ast}$), which traces crowded regions in the LMC's disk. {\em Middle Panel}: The radial profile of the spatial distribution of $\sigma_{C^\ast}$. The profile approaches a constant value in the outer disk, which indicates an intrinsic scatter to $C^\ast$ ($\sigma_{C^\ast}^{int}$) presumably due to instrumental and astrophysical effects independent of crowding. We find this constant value to be 0.04 (depicted by the black dashed line) by fitting the radial profile data with a tangent hyperbolic function. {\em Right Panel}: The spatial distribution of $\sigma_{C^\ast}^{crowd}$, which we obtain by subtracting the intrinsic scatter in $C^\ast$ from the total scatter in quadrature. $\sigma_{C^\ast}^{crowd}$ directly probes crowding in the LMC's disk.}
     \label{fig:sigma_Cstar_maps}
\end{figure*}

In the left panel of Figure \ref{fig:sigma_Cstar_maps}, we show the spatial distribution of $\sigma_{C^\ast}$, in bins of $(\Delta \xi \times \Delta \eta) = (0.02^\circ \times 0.02^\circ)$. We observe that $\sigma_{C^\ast}$ traces crowded regions in the LMC's disk. However, $C^\ast$ may have an intrinsic scatter ($\sigma_{C^\ast}^{int}$) that is related to instrumental and astrophysical\footnote{For example, for completely isolated sources, the combined BP+RP flux compared to the G flux may be slightly different from source to source because of differences in their spectral energy distributions (SEDs).} effects, but not to stellar crowding itself. We measure $\sigma_{C^\ast}^{int}$ by analyzing the spatial distribution of $C^\ast$ in the outskirts of the LMC's disk, where crowding does not significantly affect the stellar photometry. In the middle panel of Figure \ref{fig:sigma_Cstar_maps}, we show the radial profile of the spatial distribution of $\sigma_{C^\ast}$ and find that indeed the profile flattens out to a constant value at larger radii. To determine this constant value, we fit the radial profile with a tangent-hyperbolic model:

\begin{equation} \label{eq:sigma_Cstar_R}
\sigma_{C^\ast} (R) = \Sigma_0 - \Sigma_1 \tanh{\left(\frac{R - R_0}{\sigma_R} \right)}
\end{equation}

We use the python package scipy.optimize.curvefit to perform this fit. This package performs the fit in a least squares sense. We get the following values for the model parameters of equation (\ref{eq:sigma_Cstar_R}):

$\Sigma_0 = 0.14$, $\Sigma_1 = 0.10$, $R_0 = 1.19$, $\sigma_R = 2.62$

For calculating $\sigma_{C^\ast}^{int}$, we take the limit of (\ref{eq:sigma_Cstar_R}) at large radii:

\begin{equation}
    \sigma_{C^\ast}^{int} = \lim_{R\to\infty} \sigma_{C^\ast} (R) = \Sigma_0 - \Sigma_1 = 0.04
\end{equation}

We subtract the intrinsic scatter in $C^\ast$ from the overall scatter in quadrature to obtain a statistical quantity that probes crowding, which we call $\sigma_{C^\ast}^{crowd}$. We find that the dependence of $\sigma_{C^\ast}$ on other quantities like color and magnitude is less than its intrinsic scatter, and thus is not expected to significantly affect the completeness calibration. Hence, we consider just the spatial dependence.

\begin{equation} \label{eq:deconvolve}
    \sigma_{C^\ast}^{crowd} = \sqrt{\sigma_{C^\ast}^2 - (\sigma_{C^\ast}^{int})^2} 
\end{equation}

We show the spatial distribution of $\sigma_{C^\ast}^{crowd}$ in the right panel of Figure \ref{fig:sigma_Cstar_maps}.

Next, we convert the 2-D map of $\sigma_{C^\ast}^{crowd}$ to the 2-D map of completeness. For this, we refer to the calibration of O03, and fit the following curve (equation \ref{eq:completeness_cal}) to a representative sample of their Figure 14 data points, which show the relation between the completeness fraction and the error in color inference of stars due to crowding:

\begin{equation} \label{eq:completeness_cal}
f = f_0 \left[1 - \tanh{\left(\frac{\sigma_{C^\ast}^{crowd} - \sigma_0}{\sigma_s} \right)}\right]
\end{equation}

Where $f$ is the completeness fraction, and $(f_0, \sigma_0, \sigma_s)$ are the fit parameters of equation (\ref{eq:completeness_cal}). Using the python package scipy.optimize.curvefit, we obtain the following values for the fit parameters:

$$f_0 = 0.50, \sigma_0 = 0.13, \sigma_s = 0.07$$

O03 have verified the above formulation with AST simulations applied to NGC 1835 - a globular cluster in the LMC, which is an example of a highly crowded source scenario like the LMC's bar. They consider a large range of telescope spatial resolution ($\sim$ 2 dex) in their simulations, ranging from seeing limited observations from the ground, diffraction limited observations from the Hubble Space Telescope, and simulated observations from Adaptive Optics corrected Extremely Large Telescopes, and find their formulation to be robust.

\begin{figure}
    \centering
    \includegraphics[width=\columnwidth]{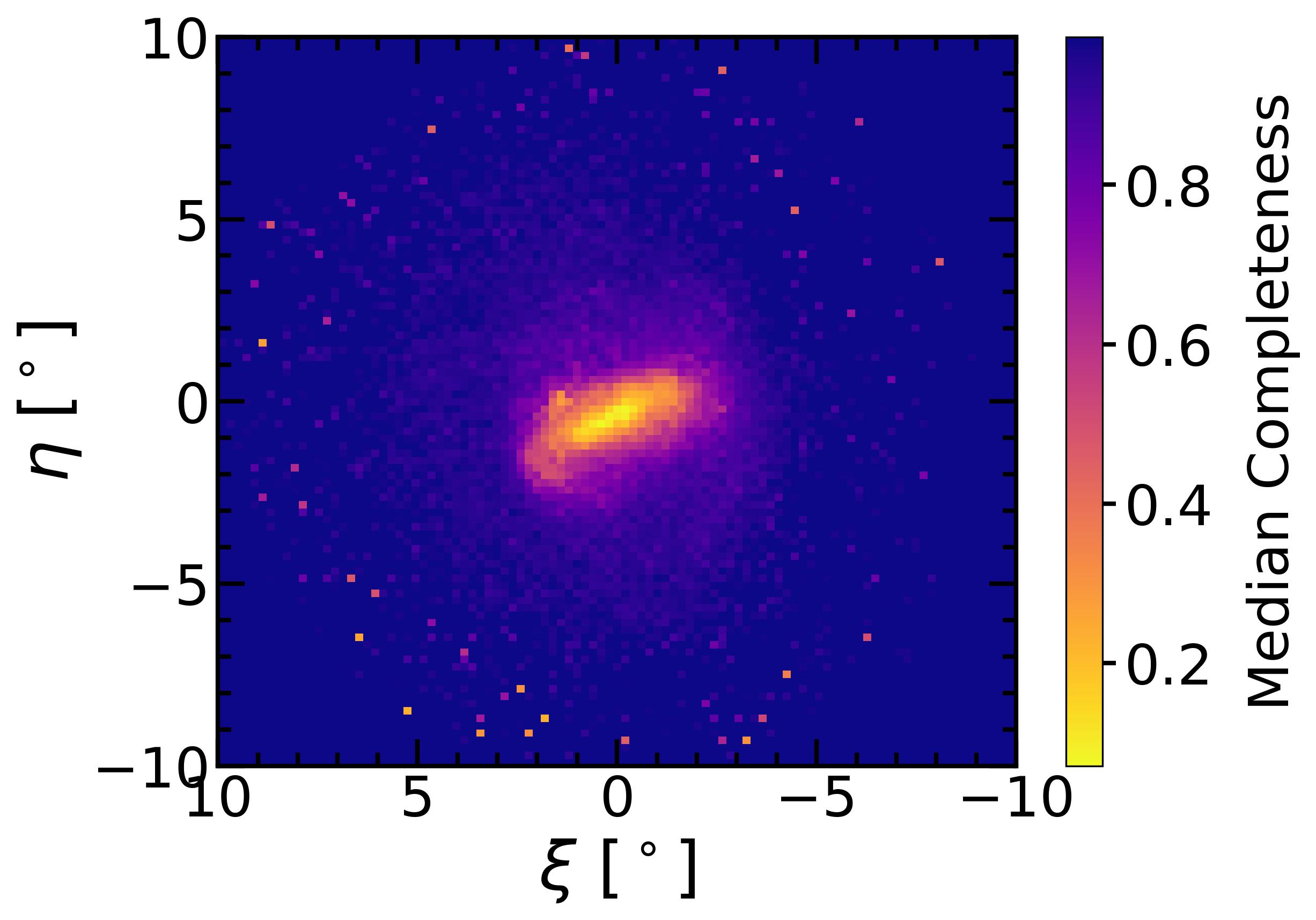}
    \caption{The completeness map of the LMC's disk derived from the spread in the Gaia BP-RP color excess. The map shows the median completeness in spatial bins of $(\Delta \xi \times \Delta \eta) = (0.02^\circ \times 0.02^\circ)$. A completeness of 1 indicates $100\%$ of the stars are being counted in that spatial bin, while a completeness of 0 indicates most of the stars are not being counted.}
    \label{fig:comp_man}
\end{figure}

\begin{figure}
    \centering
    \includegraphics[width=\columnwidth]{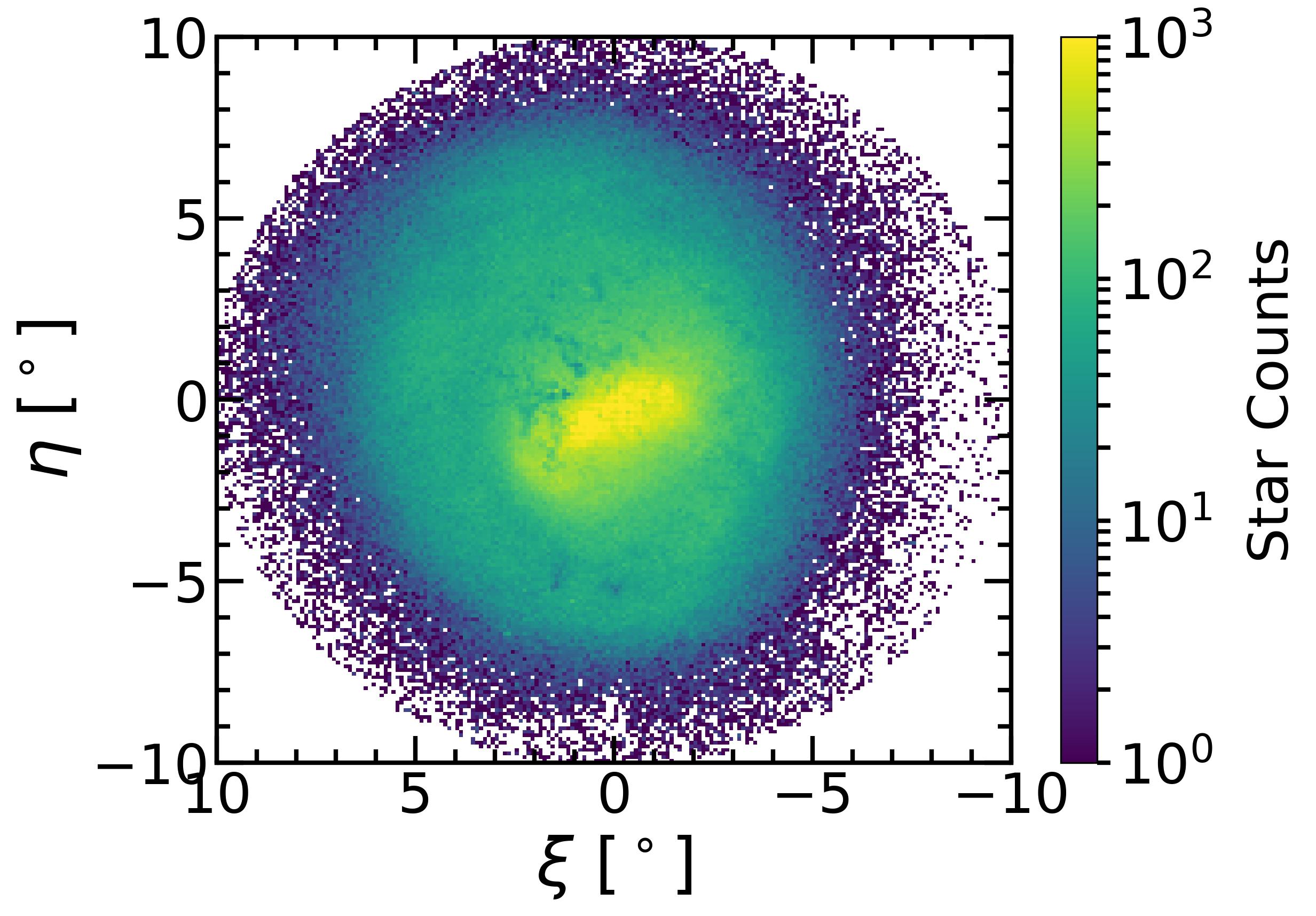}
    \caption{Completeness corrected LMC disk obtained by weighting the incomplete disk in Figure \ref{fig:cuts} ({\em right panel}) by the inverse of the completeness map (Figure \ref{fig:comp_man}). Central region becomes much more prominent after the completeness correction.}
    \label{fig:disk_weighted}
\end{figure}

Figure \ref{fig:comp_man} shows the completeness map of the LMC's disk using equation (\ref{eq:completeness_cal}). A completeness fraction of 1 in a spatial bin indicates $100\%$ of the red clump stars residing in that bin are being counted, while a completeness close to 0 indicates most of the red clump stars are not being counted. We show the spatial distribution of the counts of the red clump stars weighted by the inverse of completeness in Figure \ref{fig:disk_weighted}. This is, therefore, the completeness corrected LMC star count map. We find that after the completeness correction, the central region, including the bar, become much more pronounced compared to the incomplete LMC disk (right panel of Figure \ref{fig:cuts}). 

We find that even after the completeness correction, there exist a few patches in the LMC's disk where the star counts drop locally. We attribute these patches to regions where the dust extinction is high. Our framework will miss red clump stars that reside towards the high-end tail of the reddening distribution, even after the completeness correction, since they are omitted from our simple red clump selection box. We currently do not have a prescription to correct the counts for this effect. However, our completeness correction fully accounts for the selected red clump stars, so we do not expect the missed, highly-extincted red clump stars to significantly affect our bar analysis.

\begin{figure}
    \centering
    \includegraphics[width = \columnwidth]{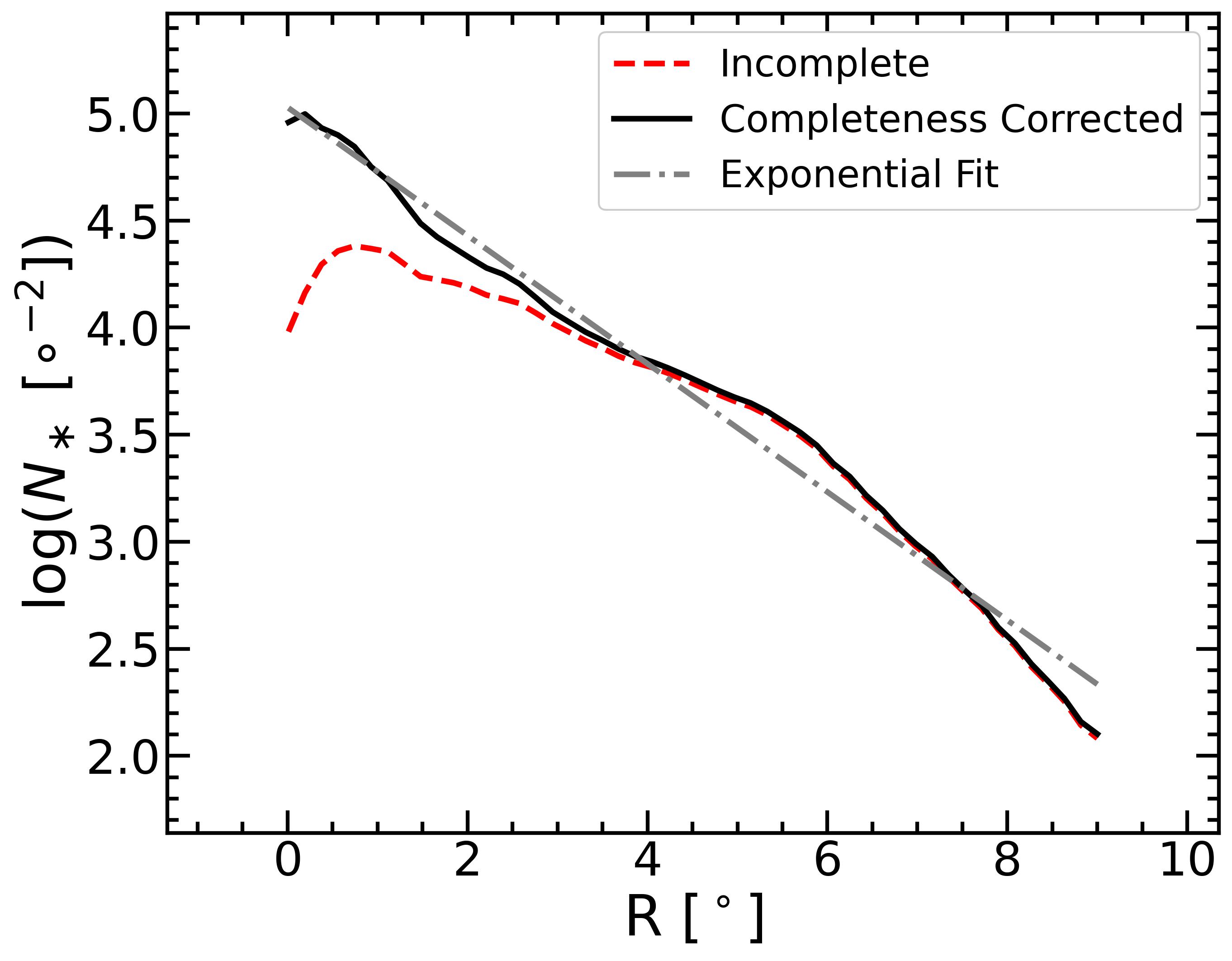}
    \caption{Red clump number density profile for the incomplete (red dashed) and the completeness corrected (black solid) LMC disks. The number density in the central regions is significantly underestimated in the incomplete disk, which is corrected for in the completeness corrected disk. The dash-dot grey line represents the best fit exponential profile to the completeness corrected data. The agreement of the data with the exponential profile in the inner disk ($R < 5^\circ$) further corroborates our completeness correction framework, where no prior information on the density profile is utilized.}
    \label{fig:dens_profile}
\end{figure}

From the completeness map we generate a completeness catalog, wherein we assign a completeness value to each star residing in a given $\xi$ - $\eta$ bin of the completeness map. We treat the inverse of the completeness as a factor that needs to be multiplied to the star counts in a given spatial bin to obtain the \enquote{true} (completeness corrected) count. In Figure \ref{fig:dens_profile}, we show the radial number density profile for red clump stars in the incomplete and completeness corrected LMC disks. We have centered both profiles on the completeness corrected center of mass (described in section \ref{sec:com}).

We find that the number density profile for the incomplete sample decreases as we go towards the inner regions, which is a crowding artifact. 

However, in the completeness corrected case, the profile is closer to an exponential disk, in accordance with expectations \citep[e.g.][]{vdM2001, Choi2018b, Frankel2024}. To illustrate this, in Figure \ref{fig:dens_profile} we over-plot the best fit exponential profile to the completeness corrected number densities:
\begin{equation}
N(R) = N_0 e^{\frac{-R}{R_s}}
\end{equation}
\noindent where $N(R)$ is the number density of red clump stars at a radius $R$, $N_0 = N(R = 0)$ and $R_s$ is the scale radius of the profile. We find $N_0 = 1.13 \times 10^5$ degree$^{-2}$ and $R_s = 1.45$ kpc. The agreement between the best fit exponential profile and the data in the inner regions of the LMC ($R < 5^\circ$) further corroborates our completeness correction framework, where no prior information on the shape of the density profile is utilized. The profiles for the two samples agree in the outer regions, where completeness is close to unity.

\cite{Arranz2024} (hereafter JA24) use Gaia selection functions \citep{CantatGaudin2023, Castro-Ginard2023} to estimate an average completeness $>$ 50\% in the LMC bar region for stars with $19 < G < 19.5$. Based on our color excess framework, we obtain a median completeness of 64\% for the red clump stars ($18.6 < G < 19.3$) within $2.5^\circ$ of the LMC's center, which is consistent with the prediction of JA24 for our red clump magnitude range. This higher completeness compared to JA24 is expected because our analysis uses brighter stars.

Note that we compute only the completeness with respect to the magnitude and color range corresponding to our red clump selection box. Our framework does not correct for other stellar populations (e.g., main sequence) outside our selection box.

\section{Fourier Decomposition of the LMC's Disk} \label{sec:fourier}
In this section, we describe the framework to extract the LMC bar properties from the completeness-corrected observations and simulations. Fourier decomposition allows us to systematically quantify the bar in the disk, and has been widely used in galaxy dynamics literature to study bar properties \citep[e.g.][]{AM2002, Lucey2023, Silva2023, Ansar2023, Cuomo2024}. 

First, we consider the surface density of the disk $\Sigma(R, \theta)$ in polar coordinates $(R, \theta)$.
We take a Fourier transform in the azimuthal coordinate as follows:
\begin{equation} \label{eq:alpha}
    \alpha_m(R) = \frac{1}{\pi}\int_0^{2\pi} \Sigma (R, \theta) \cos(m\theta)d\theta, \qquad m = 0, 1, 2, ...
\end{equation}
\noindent where the integer $m$ quantifies the spatial frequency harmonic, and,
\begin{equation} \label{eq:beta}
    \beta_m(R) = \frac{1}{\pi}\int_0^{2\pi} \Sigma (R, \theta) \sin(m\theta)d\theta, \qquad m = 1, 2, 3, ...
\end{equation}
The normalized Fourier amplitude is given by:
\begin{equation} \label{eq:A}
    A_m = \frac{\sqrt{\alpha_m^2 + \beta_m^2}}{A_0}, \qquad m = 1, 2, 3, ...
\end{equation}
\noindent where $A_0 = \sqrt{\alpha_0^2 + \beta_0^2}$.
The Fourier phase is given by:
\begin{equation} \label{eq:phi}
    \Phi_m = \frac{1}{m}\arctan\left(\frac{\beta_m}{\alpha_m}\right) \qquad m = 1, 2, 3, ...
\end{equation}

In Gaia datasets and in N-body simulations we have the positions of individual stars and star particles respectively, so we bin the disk in radial annuli and discretize the azimuthal Fourier transform (equations \ref{eq:alpha} and \ref{eq:beta}):

\begin{equation} \label{eq:alpha_dis}
\alpha_m = \frac{\sum_i M_i w_i\cos(m \phi_i)}{N},
\end{equation}

\begin{equation} \label{eq:beta_dis}
\beta_m = \frac{\sum_i M_i w_i\sin(m \phi_i)}{N},
\end{equation}

\begin{equation} \label{eq:A_dis}
A_m = N \frac{\sqrt{\alpha_m^2 + \beta_m^2}}{\sum_i M_i},
\end{equation}

\noindent where $M_i$ is the mass of each star/star particle in a radial bin and $N = \sum_i i$ is the total number of stars/star particles in the radial bin. $w_i$ is a weight associated with each star, which is calculated as the inverse of the completeness fraction at the location of the star. For simulations, we will set this weight to 1. We shall assume that each red clump star has the same mass. This is a reasonable assumption since red clump stars have a narrow distribution of masses \citep{Castellani2000}.

\subsection{Measuring Bar Parameters} \label{sec:bar_params_fourier}
Fourier coefficients ($\alpha_m$ and $\beta_m$) corresponding to a frequency harmonic $m$ have a periodicity of $\frac{2\pi}{m}$. The bar is the dominant bi-symmetric structure in the inner disk of a galaxy. Thus, for characterizing the bar, we use the $m = 2$ spatial frequency harmonic. 

The Fourier amplitude $A_2$ characterizes the strength of the bar, which is the contribution of the bar to the gravitational potential of the disk \citep{Guo2019}. 
In general, $A_2(R)$ will rise as we move outwards from the disk center, attain a peak, and then drop sharply around or after the end of the bar. In literature, galaxies with a peak $A_2 < 0.20$ are considered unbarred or weakly barred, whereas galaxies with a peak $A_2 > 0.20$ are considered barred \citep[e.g.][]{Ansar2023}.

The Fourier phase $\Phi_2$ characterizes the phase angle of the bi-symmetric component, which corresponds to the position angle of the bar. $\Phi_2 (R)$ will be nearly constant in the bar region, and will start deviating around or after the end of the bar. 

The radius at which $A_2(R)$ drops significantly, or where $\Phi_2(R)$ significantly deviates from a constant, can be used to identify the end of the bar and thereby define the bar radius. Three metrics have been commonly used in literature to determine the bar radius:

\begin{itemize}
    \item The radius at which $A_2$ drops to $0.15$ after attaining a peak. We take this definition from \cite{Lucey2023}, and hereafter refer to it as the L23 metric.
    \item The radius at which $A_2$ drops to $50$\% of its maximum value. We take this definition from \cite{AM2002}, and hereafter refer to it as the AM1 method.
    \item The radius at which $|\Phi_2|$ deviates by more than $10^\circ$ from a constant. Taken from \cite{AM2002}. Hereafter referred to as the AM2 method.
\end{itemize}

The threshold values for identifying the end of the bar in the metrics above do not have a rigorous mathematical justification, but are commonly used in literature. The three metrics are not expected to give identical measures of the bar radius, however, typically they agree within a kpc \citep[e.g.][]{AM2002, Ghosh2024}. The metric based on the bar phase (AM2) is in general preferred over the other metrics since it clearly indicates if there is a significant contribution from other bi-symmetric structures like a 2-armed spiral in the Fourier profile of the inner disk \citep{Ghosh2024, Cuomo2024}. When comparing the bar radius across different observations or simulations, it is important to use a consistent metric.

\subsection{Centering the LMC's Disk} \label{sec:com}
The Fourier coefficients are sensitive to the choice of the center of mass. A shift in centering will cause a shift in the Fourier phase, as a consequence of the Fourier shift theorem. Thus, it is important to identify the correct center of mass for the distribution of stars. For this, we use an iterative shrinking sphere method following \cite{Power2003}. We first cover the two dimensional projection of the disk with a sphere of radius $10^\circ$ and compute the center of mass of the particles inside that sphere. Then, we shrink the sphere by $30\%$ and compute the center of mass of the particles that remain inside the shrunken sphere. We repeat this shrinking process until the center of mass of the particles that remain inside the sphere converges (does not differ by more than 0.01 kpc). 

This method is particularly robust when there are several asymmetric features in the outer parts of the disk which can otherwise potentially bias the center of mass if inferred from a simple mass weighted average. Moreover, it has been inferred using simulations that this method converges to the center of mass obtained by weighting the position of each star with the gravitational potential at that position \citep{Power2003}. Thus the center inferred from this method is also the stellar dynamical center of mass. This method has been frequently applied in simulations \cite[e.g.][]{GC2019}. With the completeness correction framework, we can apply this method to star count maps in observations to correctly center the Fourier profiles.

After the completeness correction, we obtain the following stellar dynamical center of mass of the LMC's disk based on red clump stars:

$$(\rm{RA}, \: \rm{DEC}) = (80.27^\circ, -69.65^\circ)$$

In Table \ref{tab:com}, we report our stellar dynamical center of mass estimate for the LMC for both incomplete and completeness corrected samples, along with other center of mass values reported in literature. In Figure \ref{fig:bar_com_phase}, we show the location of the various center of masses. We find that incompleteness significantly affects the inference of the stellar dynamical center of mass.

\begin{table*}
    \centering
    \caption{Various center of masses for the LMC's disk}
    \begin{tabular}{c c c c}
    \hline
    \hline
     COM Type & Reference  & Value (RA, DEC) $[^\circ]$ & Separation [$^\circ$]  \\
     \hline
     \textbf{Stellar Dynamical (Completeness Corrected)} & \textbf{
     This Work} &\textbf{(80.27, -69.65)} & 0 \\
     Stellar Dynamical (Incomplete) & This work  & (78.89, -70.04) & 0.62 \\
     Stellar Kinematic & C22 & (80.44, -69.27) & 0.38 \\
     Photometric & \cite{vdM2001} & (81.28, -69.78) & 0.37 \\
     HI Kinematic & \cite{Kim1998} & (79.35, -69.03) & 0.7 \\
     \hline
    \end{tabular}
    \tablenotetext{}{\textbf{Note:} In this work, we determine a stellar dynamical center of mass by applying the iterative shrinking sphere algorithm to the spatial distribution of red clump counts in Gaia DR3. We report the center of mass for both completeness corrected and incomplete samples. We observe that incompleteness significantly affects the center of mass inference. We also report the values of the stellar kinematic center of mass, the photometric center of mass and the HI kinematic center of mass from literature. In the last column of the table, we find the on-sky separation between the center of mass calculated in this work and the other center of masses reported. We find that the HI kinematic center is separated from the stellar dynamical center by almost a degree.}
    \label{tab:com}
\end{table*}

\section{The Geometry of the LMC's Bar} \label{sec:lmc_bar}

In this section, we apply the techniques discussed in the previous section to measure the LMC bar's properties from our complete red clump star catalog. We shall analyze the bar properties in the LMC in-plane cartesian coordinate system, to obtain parameters in physical units.

\subsection{Bar Position Angle} \label{sec:pos_angle}

\begin{figure}
    \centering
    \includegraphics[width = \columnwidth]{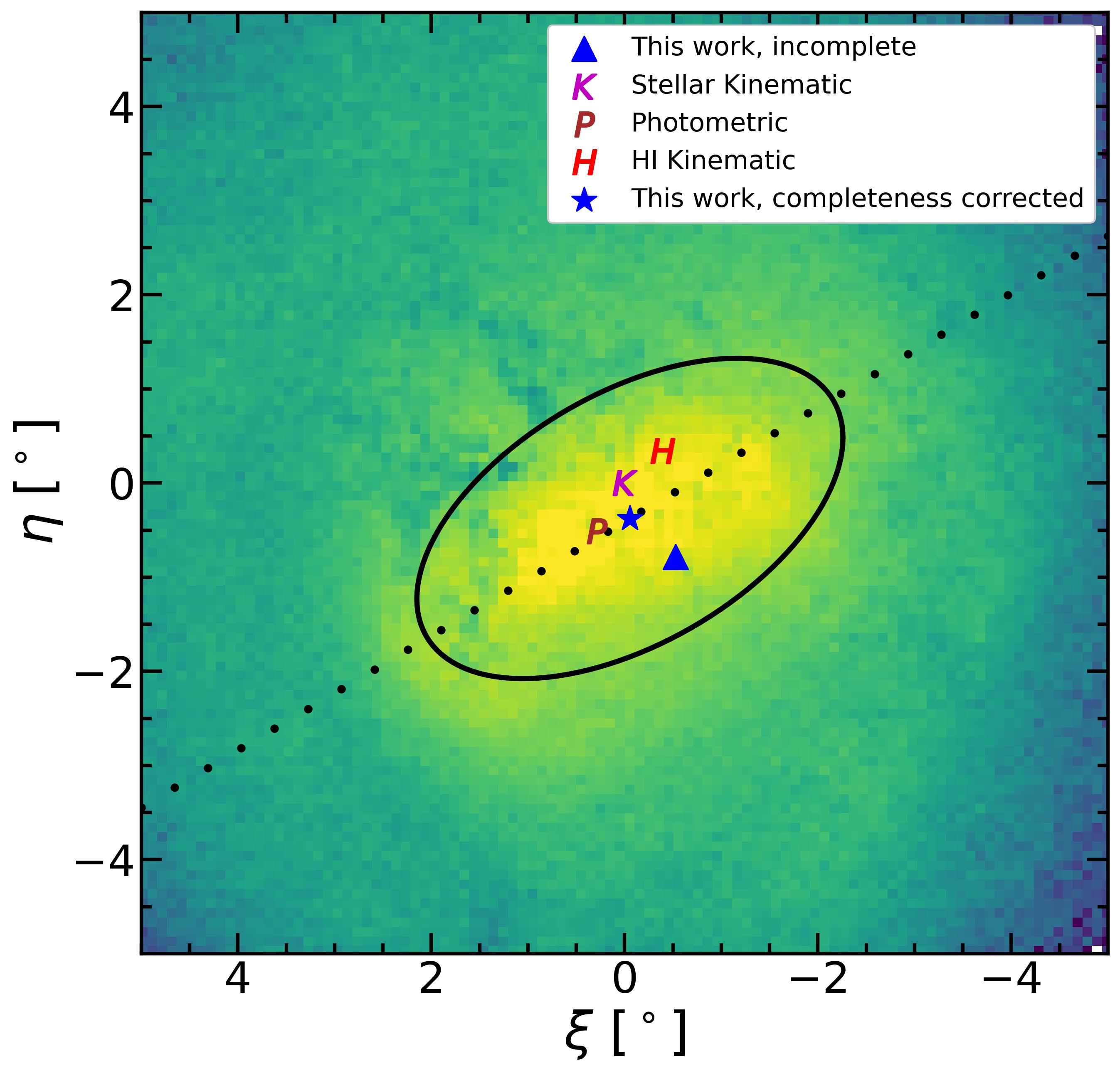}
    \caption{Various estimates for the center of mass of the LMC's disk. We depict the stellar dynamical center of mass for the completeness corrected and incomplete samples using the blue star and blue triangle respectively. The magenta \enquote{K} depicts the LMC stellar kinematic center found by C22, the brown \enquote{P} the LMC photometric center found by \cite{vdM2001}, and the red \enquote{H} the HI kinematic center found by \cite{Kim1998}. We find that incompleteness significantly affects the determination of the stellar dynamical center of mass. The dotted black line depicts the position angle of the LMC's bar calculated using Fourier decomposition. The remarkable agreement between the inferred phase angle and the visual expectation based on the location of the bar (shown by the black ellipse) confirms the reliability of the Fourier decomposition technique. In these coordinates, north is towards the top and east is towards the left.}
    \label{fig:bar_com_phase}
\end{figure}

The bar position angle is the $m = 2$ phase of the Fourier series (equation \ref{eq:phi}), measured using the appropriate convention \footnote{We use the convention of \cite{vdM2001} where the bar position angle is measured from the north to the east in a counter-clockwise sense. In our tangent plane coordinate system (Figure \ref{fig:bar_com_phase}), north is towards the top (positive $\eta$ axis) and east is towards the left. So the position angle is measured from top to left.}. We construct an aperture on the disk centered at the dynamical center of mass determined in section \ref{sec:com}, and compute the phase by performing a Fourier decomposition of stars within that aperture. We calculate the phase as a function of the aperture size, and find that it remains constant for apertures larger than $\sim 1$ kpc, indicating the dominance of the bar. We take the average of the inferred phase between 2 kpc and 4 kpc.

To obtain an error estimate on the bar phase, we generate bootstrapped realizations of the data by resampling the positions of the stars 1000 times, and calculate the phase for each realization. We find that the distribution of phases in the bootstrapped realizations is approximately Gaussian. We use the standard deviation of the bootstrapped phases as our error metric. We report the position angle of the LMC's bar to be:

$$\rm{PA}_{\rm{bar}} = 121.26^\circ \pm 0.21^\circ$$

We illustrate this position angle in Figure \ref{fig:bar_com_phase}. The position angle obtained from Fourier decomposition aligns very well with the visual expectation, affirming the reliability of our method. Our measured bar position angle implies that the LMC's bar is misaligned with the line of nodes of the disk ($\sim 150^\circ$) \citep{Choi2018a} by around 30$^\circ$, which is consistent with JA24. 

We repeat the above analysis for the incomplete sample of red clump stars, and find a position angle of $106.49^\circ \pm 0.34^\circ$, which is very different from the value obtained with the completeness corrected sample. This re-iterates that the completeness correction is a necessary step for studying the bar geometry.

\subsection{Bar Strength and Radius} \label{sec:bar_strength_rad}

We compute the bi-symmetric Fourier amplitude $A_2$ (equation \ref{eq:A_dis}) as a function of radial distance. First, we align the bar with the x-axis of the coordinate system (a phase of 0$^\circ$) by rotating the disk according to the position angle measured in section \ref{sec:pos_angle}. Then, we bin the LMC's disk in radial annuli of width 0.1 kpc, and compute $A_2$ in each annulus. 

\begin{figure*}
    \centering
    \includegraphics[width = \textwidth]{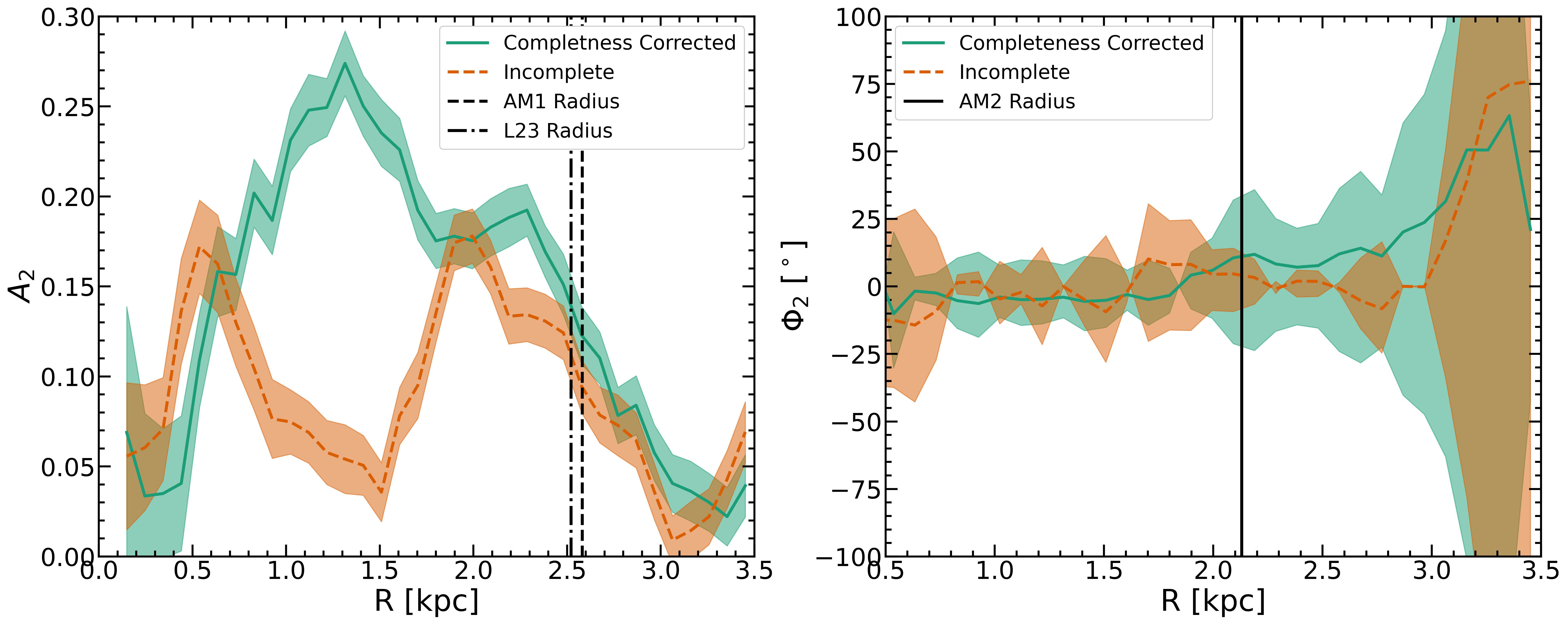}
    \caption{Radial profile of the $m = 2$ Fourier amplitude $A_2$ (left panel) and the Fourier phase $\Phi_2$ (right panel) for the completeness corrected (blue, solid lines) and the incomplete (orange, dashed lines) samples of red clump stars in the LMC disk. The shaded regions denote the 3-$\sigma$ spread of the bootstrapped realizations. {\it Left Panel:} The increase in the Fourier amplitude with radius in the inner disk, the attainment of a peak, and the subsequent decrease in amplitude are concrete signatures of the bar. Incompleteness significantly underestimates the bar amplitude. {\it Right Panel:} The phase angle remains roughly constant along the bar, and deviates near the end of the bar. We mask $R < 0.5$ kpc since the radial annuli used for binning lies completely inside the bar, making the $m = 2$ phase ambiguous. The bar radius obtained from the AM1, AM2 and L23 metrics is denoted by the black dashed line, black solid line and the black dash-dot line respectively. The AM2 metric, which is based on the $m = 2$ phase, is our favored measure of the bar radius as explained in the text.}
    \label{fig:bar_fourier}
\end{figure*}

The radial profile of $A_2$ and the bi-symmetric phase $\Phi_2$ for the completeness corrected and incomplete samples is shown in Figure \ref{fig:bar_fourier}. The shaded band represents the 3-$\sigma$ spread of 1000 bootstrapped realizations. 

For the completeness corrected sample, we find that $A_2$ increases with radius, attains a peak and then decreases, in accordance with expectations (section \ref{sec:bar_params_fourier}). This is a concrete signature of the bar \citep{AM2002}. The phase angle remains constant in the bar region up to about $\approx 1.75$ kpc, and then starts deviating around the end of the bar. 

For the incomplete sample, the presence of a bar from the $A_2$ profiles is not clear. The bar amplitude is significantly underestimated. The profiles for the two samples agree in the outer regions of the disk where completeness is close to unity. 

From the completeness corrected sample, we find the peak $A_2$, which quantifies the bar strength ($S_{bar}$) of:
$$S_{bar} = 0.27 \pm 0.01$$
where the error is $1-\sigma$ of the bootstrapped realizations. This is above the threshold of $0.20$ used to ascertain the presence of a bar (section \ref{sec:bar_params_fourier}), which indicates that the LMC's disk has a significant bar perturbation. For the incomplete sample, we find a peak bar strength of $S_{bar} = 0.18 \pm 0.01$, which indicates that it is hard to ascertain the presence of a bar with incompleteness.

We use three metrics to determine the bar radius as outlined in section \ref{sec:bar_params_fourier}. Obtaining an error estimate on the bar radius is non-trivial and cannot be done with bootstrapping alone. We find that the inferred bar radius for each bootstrapped realization is very similar, and does not capture the error. This is because the sampling of the stars in the Fourier profile is not the dominant source of uncertainty in bar radius. In all three metrics we use to determine the radius, the inferred value is based on the radius where the Fourier amplitude or phase achieves a particular absolute or relative value. Thus, it is the radial binning prescription that is expected to dominate the uncertainty in the bar radius. To get an estimate of this uncertainty, we generate the Fourier profiles in a variety of different bins ranging from 0.05 kpc to 0.35 kpc and determine the radius for each profile. Then, we compute the resulting distribution of the inferred bar radius. For the completeness corrected sample, we find the bar radius corresponding to the three different metrics:

$$\rm{R}_{\rm{bar, AM1}} = 2.58^{+0.04}_{-0.05} \: \: \rm{kpc}$$ 
$$\mathbf{R_{bar, AM2} = 2.13^{+0.03}_{-0.04} \: \: kpc}$$
$$\rm{R}_{\rm{bar, L23}} = 2.52^{+0.04}_{-0.04} \: \: \rm{kpc}$$

The errors depict the 16-84 percentile spread of the distribution of the bar radius.

We prefer the AM2 metric, which is based on the $m = 2$ phase, as it is the popular choice in literature (section \ref{sec:bar_params_fourier}). We have shown the bar radius inferred from the other two metrics for the purpose of completeness.

We repeat the above analysis for the incomplete sample, and list the resulting measurements of the bar radius in Table \ref{tab:bar_props}. Incompleteness not only biases the bar radius, but leads to significantly higher uncertainties.

We do not expect the bar offset to significantly affect our analysis, since we have ensured accurate centering on the bar, as described in section \ref{sec:com}. In addition to the offset, it is well known that the LMC's bar is also lopsided (more mass being present on one end of the bar relative to the other). The lopsidedness is likely due to a spiral arm attached to the western end of the bar \citep{vdM2001, JD2016}. Bar Lopsidedness manifests as odd-m terms of the Fourier series \citep{Pardy2016, Lokas2021}. In the completeness corrected sample, we find a significant $A_1$ component in the bar region after centering on the bar, confirming its lopsided nature. Since the odd-m terms of the Fourier series (equations \ref{eq:alpha_dis} and \ref{eq:beta_dis}) are orthogonal to the even-m terms, we do not expect the bar lopsidedness to significantly affect our measurements of bar strength and radius from the m = 2 term. Indeed, the m = 2 term has been successfully used to measure bar properties in interacting/distorted barred galaxy simulations which show asymmetries such as offset and lopsidedness, after accurate centering \citep[e.g.][]{Gerin1990, Berentzen2003}.

\subsection{Bar Axis-Ratio and Offset}

The bar-axis ratio quantifies the thickness of the bar. It is defined as the ratio of the semi-minor to the semi-major axis of the bar $\left(\frac{b}{a}\right)_{bar}$.

We first align the bar with the x-axis of the coordinate system using the position angle derived in section \ref{sec:pos_angle}. We then use the python package \textit{photutils} to fit elliptical iso-contours to the spatial distribution of red clump stars in the LMC's disk. This package performs the fit in an iterative manner following \cite{Jed1987}. We define the bar ellipse as the fitted iso-ellipse whose semi-major axis equals the bar radius as inferred from the AM2 method in section \ref{sec:bar_strength_rad}. We define the axis ratio of the bar as the axis ratio of this bar ellipse. We compute the error in the axis ratio by varying the semi-major axis of the bar ellipse within the error bounds of the bar radius.

For the completeness corrected sample, we find an axis ratio of:
$$\left(\frac{b}{a}\right)_{bar} = 0.54 \pm 0.03$$

To quantify the offset of the bar with respect to the outer disk, we compute the separation between the center of the bar ellipse and the center of the iso-ellipse of the outer LMC disk. We define the outer disk ellipse as the iso-ellipse whose semi-major axis coincides with the radius where the number density profile of the disk (Figure \ref{fig:dens_profile}) drops by a factor of $100$ relative to the central value. For the completeness corrected red clump disk, this outer disk radius is 4.92 kpc. We find the bar offset for the completeness corrected sample to be:

$$\Delta_{bar} = 0.76 \pm 0.01 \: \: \rm{kpc},$$
where the error is computed by varying the bar ellipse within the bounds of the inferred bar radius.

\begin{figure*}
    \includegraphics[width=0.48\textwidth]{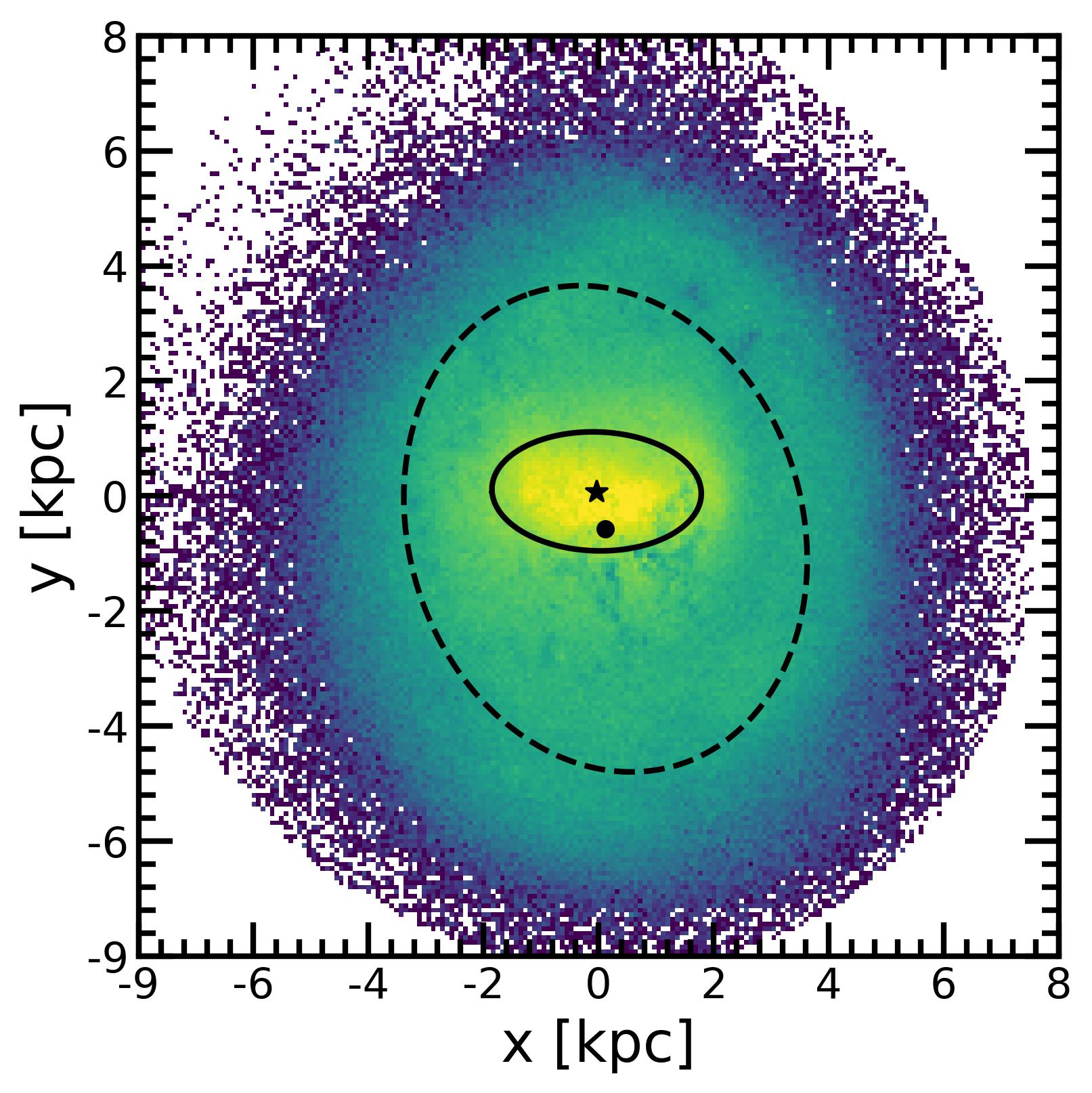}
    \includegraphics[width=0.48\textwidth]{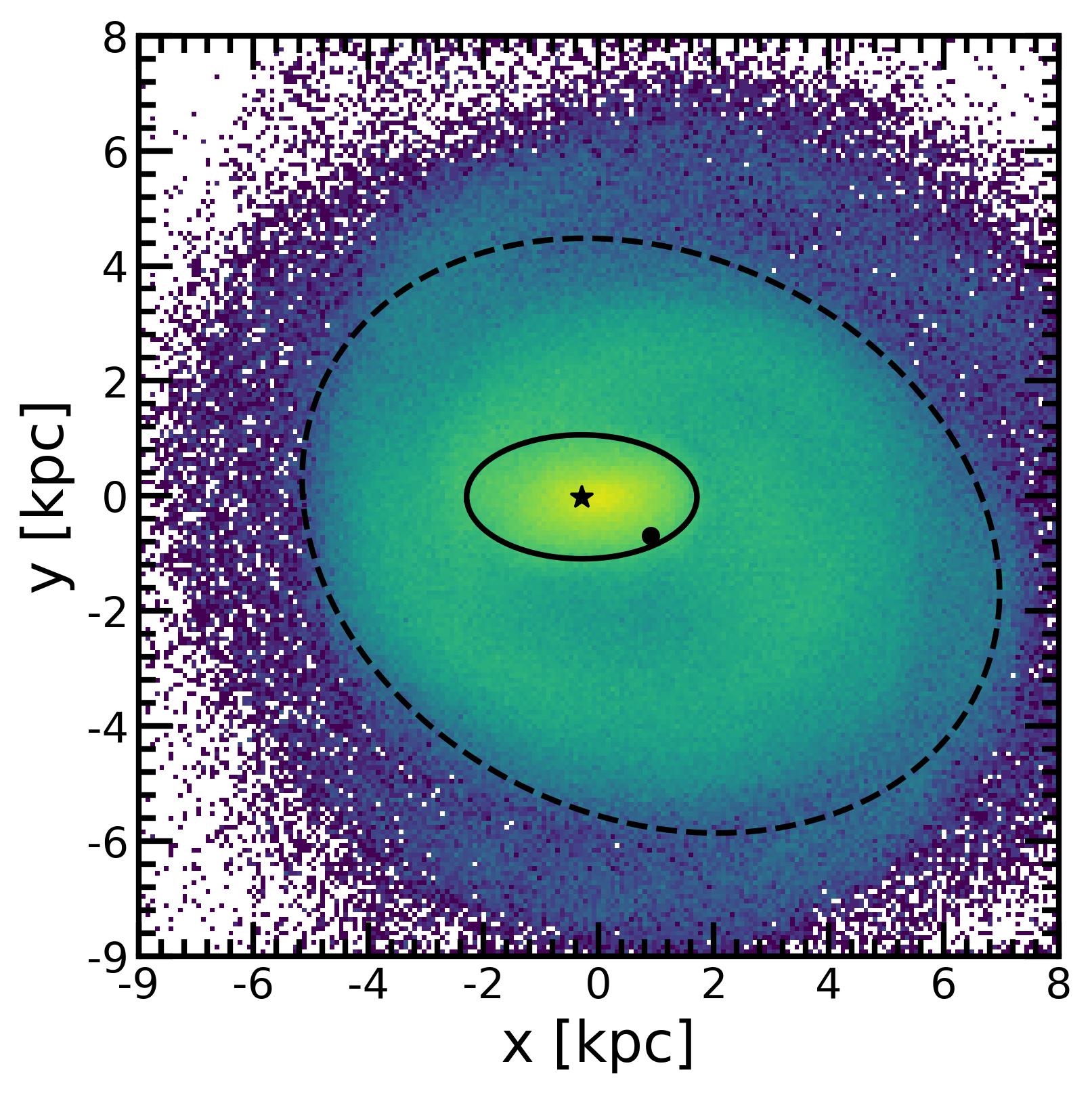}
    \caption{Illustration of the offset bar of the LMC. The {\em left panel} shows the Gaia DR3 red clump disk in the LMC in-plane cartesian coordinates, corrected for completeness. The {\em right panel} shows the LMC disk in the B12 simulation, where the SMC has undergone a recent direct collision with the LMC. The solid ellipse is the iso-contour corresponding to the bar, and the dashed ellipse is the iso-contour corresponding to the outer disk. The star denotes the center of the iso-ellipse of the bar, and the filled circle denotes the center of the iso-ellipse of the outer disk. The two centers do not coincide, which indicates an offset bar. The offset is seen in the simulated disk as well, however, it is larger relative to the observations.}
    \label{fig:bar_ellip}
\end{figure*}

We show the outer disk ellipse, the bar ellipse and their respective centers for the completeness corrected sample in the left panel of Figure \ref{fig:bar_ellip}. We repeat the analysis of the axis-ratio and offset for the incomplete sample and report our measurements in Table \ref{tab:bar_props}.

\begin{table*}
    \centering
    \caption{Our measurements for the LMC bar's geometric parameters}
    \begin{tabular}{c c c c c c c c}
    \hline
    \hline
     Dataset & \multicolumn{3}{c}{Radius [kpc]} & Position Angle & Strength & Axis Ratio & Offset\\ 
     & AM1 & AM2 & AM3 & [$^\circ$] & & & [kpc] \\
     \hline
     \textbf{Gaia DR3 Complete} & $2.58^{+0.04}_{-0.05}$ & $\mathbf{2.13^{+0.03}_{-0.04}}$ & $2.52^{+0.04}_{-0.04}$ & $\mathbf{121.26 \pm 0.21}$ & $\mathbf{0.27 \pm 0.01}$ & $\mathbf{0.54 \pm 0.03}$ & $\mathbf{0.76 \pm 0.01}$\\
     Gaia DR3 Incomplete & $1.87^{+0.81}_{-1.01}$ & $2.08^{+1.07}_{-1.42}$ & $1.52^{+0.68}_{-0.84}$ & $106.49 \pm 0.34$ & $0.18 \pm 0.01$ & $0.68 \pm 0.10$& $0.86 \pm 0.17$\\
     Besla+2012 Simulation & $2.44^{+0.06}_{-0.05}$ & $2.18^{+0.04}_{-0.05}$ & $2.17^{+0.05}_{-0.05}$ & $157.19 \pm 0.20$ & $0.25 \pm 0.01$ & $0.54 \pm 0.01 $ & $1.50 \pm 0.05$\\
     \hline
    \end{tabular}
    \tablenotetext{}{\textbf{Note:} We report the values for the completeness corrected red clump disk, the incomplete red clump disk, as well as the disk in B12 simulation where the SMC has undergone a recent direct collision with the LMC. Incompleteness not only biases the measurement of the bar properties, but also induces significantly larger uncertainties. We notice a general agreement between the simulations and observations for most of the bar parameters, which suggests consistency with a scenario where the SMC has undergone a recent direct collision with the LMC. We indicate our preferred measurements for the LMC's bar in bold-face.}
    \label{tab:bar_props}
\end{table*}

We compile all the measurements of this section, for both the completeness corrected and incomplete samples in Table \ref{tab:bar_props}. We compare our measurements of the bar parameters with a subset of literature values reported in Table \ref{tab:lit_bar} in section \ref{sec:lit_comp} and with a numerical simulation of the LMC-SMC-MW interaction history in section \ref{sec:sim}. In section \ref{sec:context}, we shall place our measurements in context with the population of barred galaxies in the local universe.

As mentioned in Section \ref{sec:intro}, the LMC’s bar is likely in a different plane relative to the overall LMC disk \citep{SS2013, Choi2018a}. A different inclination of the bar region with respect to the outer disk can cause a projection effect in our estimate of the bar length and the bar offset. However, we expect this effect to be small, since the average tilt of the LMC’s bar plane relative to the disk plane is around 10$^\circ$, which corresponds to a projection bias factor of $\cos(10^\circ) = 0.98$, which is very close to unity. Thus, the expected projection bias is within our errors for the bar measurements. 
In addition, the LMC’s inner regions might have a different position angle relative to the LMC’s outer regions \citep{Choi2018a, Dhanush2024}. However, the Fourier decomposition technique adopted in this work averages over the azimuthal coordinate. Thus the global position angle at which we choose to orient our coordinate system will not significantly affect our results as long as we ensure that the position angle is constant in the bar region.

\section{Discussion} \label{sec:discussion}

We have successfully corrected the observed LMC's stellar disk for completeness and derived the geometric parameters of its bar using the framework of Fourier decomposition. Next, we compare our measured bar parameters with a subset of previous measurements in literature as reported in Table \ref{tab:lit_bar}. 

\subsection{Comparison of our Measured Bar Parameters with Literature} \label{sec:lit_comp}

We find that the LMC bar radius measurement for the completeness corrected sample ($2.13^{+0.03}_{-0.04}$ kpc) is close to the measurement of JA24 (2.3 kpc). JA24 used the dynamical model by \cite{Dehnen2023} based on the stellar continuity equation, which yields both the bar pattern speed as well as the bar radius. This dynamical model identifies the bar as a collection of stars with a roughly constant bi-symmetric phase, which is similar to our approach in principle. We could not find error estimates in JA24, thus it is difficult to compare their measurement with ours in a statistical sense. However, given the similarities between our methods, it is reasonable that the two measurements are within 0.5 kpc. Also, JA24 estimated the amplitude of the m = 2 component of the velocity field in the LMC's central region. The peak radius of the velocity m = 2 amplitude (Figure 7 of JA24) is consistent with the peak radius of the stellar density m = 2 amplitude (Figure \ref{fig:bar_fourier}) within 0.5 kpc. Our bar radius measurement is also consistent with \cite{Kacharov2024}, who infer a radius of 2.2 kpc with Schwarzschild Orbit Modeling \citep{Schwarzschild1979} of the LMC's inner regions.  

Our bar radius measurement is significantly different from the measurement of \cite{Choi2018b} (3 kpc; hereafter C18b). C18b used a combined bar $+$ disk model to make their measurements. The mass profile of their bar component is modeled using a modified Ferrer profile \citep{Laurikainen2007, Peng2010}, and its boxy/disky nature is quantified using the generalized ellipse of \cite{Athanassoula1990}. Their disk component is modeled with an exponential profile. C18b mention that this model tends to choose the upper end of the prior for the bar radius. Thus, we anticipate that it would be challenging to obtain the true bar radius from this model because of the bias introduced by the prior. \cite{ZhaoEvans2000} also utilized a combined bar $+$ disk model, but they modeled the bar with an exclusively boxy mass profile. We anticipate their method has challenges similar to that faced by the C18b model, where a boxy prior assumption biases the model to more compact stellar distributions yielding a smaller bar radius (1.5 kpc). The measurement of \cite{deVFreeman72} (1.5 kpc) is also lower than our measurement. However, their work is based on fitting isophotes to images of the LMC, and light can be a biased tracer of stellar mass (as we have shown).

We find the LMC bar's position angle measurement (121.26$^\circ$ $\pm$ 0.21$^\circ$) to be consistent with previous estimates like \cite{ZhaoEvans2000, vdM2001} (see Table \ref{tab:lit_bar}). However, our position angle is significantly different from C18b (154.18$^\circ$). We attribute this difference to another subtlety of the combined bar $+$ disk model. They mention that when the radius of the bar reaches the maximum value allowed by the prior, the model tries to fit the surrounding stars by becoming boxier. Because of this, we anticipate that the model is including a part of the disk stars in the bar, which biases the inferred position angle of the bar. 

We find the LMC bar's axis ratio measurement (0.54 $\pm$ 0.03) to be larger than the estimates of \cite{deVFreeman72, ZhaoEvans2000, Choi2018b}, and on the higher side of the range described by \cite{vdM2001} (0.33 - 0.57). This indicates that the LMC bar might be thicker than previously thought. 

We find the LMC bar's offset (0.76 $\pm$ 0.01 kpc) to be larger compared to the estimate of \cite{vdM2001} (0.4 kpc), but closer to the estimate of \cite{ZhaoEvans2000} (0.87 kpc). Accurately measuring the bar offset is important for the dynamical modelling of the LMC's bar in presence of the SMC's interactions (see next section). 
\\
\\
Now, we are in a position to obtain insights into the evolutionary history of the LMC through its bar properties, in the context of its interactions with the SMC and the Milky Way (MW). We also compare the LMC bar's properties with bars in other galaxies to place the LMC in context with the population of barred galaxies in the local universe. 

\subsection{Comparison With a Hydrodynamic Model of a Recent LMC-SMC Collision} \label{sec:sim}

To obtain insights on the evolutionary history of the LMC through its measured bar properties, we compare our measurements with the hydrodynamic model of \cite{Besla2012} (hereafter B12). B12 modeled the interaction history of the LMC and SMC over the past 6-7 Gyr, including a MW infall for the past 1 Gyr. B12 presented two models - Model 1 and Model 2. In Model 1, the SMC and LMC remain far from each other, with the closest passage being $\ge 20$ kpc. In Model 2, the SMC undergoes a direct collision with the LMC's disk (impact parameter $\sim 2$ kpc) around $100$ Myr ago. Model 2 reproduces the observed structure and kinematics of the LMC significantly better compared to Model 1 \citep{Besla2016, Choi2018a, Zivick2019, Choi2022}, thus favoring a direct collision scenario. Moreover, the bar in Model 1 is not consistent with observations. It does not show any offset with respect to the disk and does have a prominent signature in gas. Hence, we compare our measurements to B12 Model 2. Below, we provide a brief summary of the B12 simulation setup. 

The initial live dark matter halos of the LMC and SMC are modeled with Hernquist profiles, having a mass of $1.8 \times 10^{11}$ M$_\odot$ and $2.1 \times 10^{10}$ M$_\odot$ respectively and a resolution of $1.8 \times 10^6$ M$_\odot$ and $2.1 \times 10^6$ M$_\odot$ respectively. Their initial stellar disks are modeled with exponential profiles having a stellar mass of $2.5 \times 10^{9}$ M$_\odot$ and $2.6 \times 10^{8}$ M$_\odot$ respectively with a resolution per particle of $2500$ M$_\odot$ and $2600$ M$_\odot$ respectively. Their initial gas disks are modeled with exponential profiles having a gas mass of $1.1 \times 10^9$ M$_\odot$ and $7.9 \times 10^8$ M$_\odot$ respectively with a resolution per particle of $3667$ M$_\odot$ and $2633$ M$_\odot$ respectively.

The LMC and SMC are allowed to interact with each other on an eccentric and decaying orbit. They were introduced in a static Milky Way like potential over the past 1 Gyr. The \enquote{present day} in the simulations was determined by matching the LMC's position and velocity to observed values. The bar is clearly visible in the LMC as a gaseous structure prior to the collision with the SMC, which occurred around $100$ Myr before the \enquote{present day}. 

\begin{figure*}
    \centering
    \includegraphics[width = \textwidth]{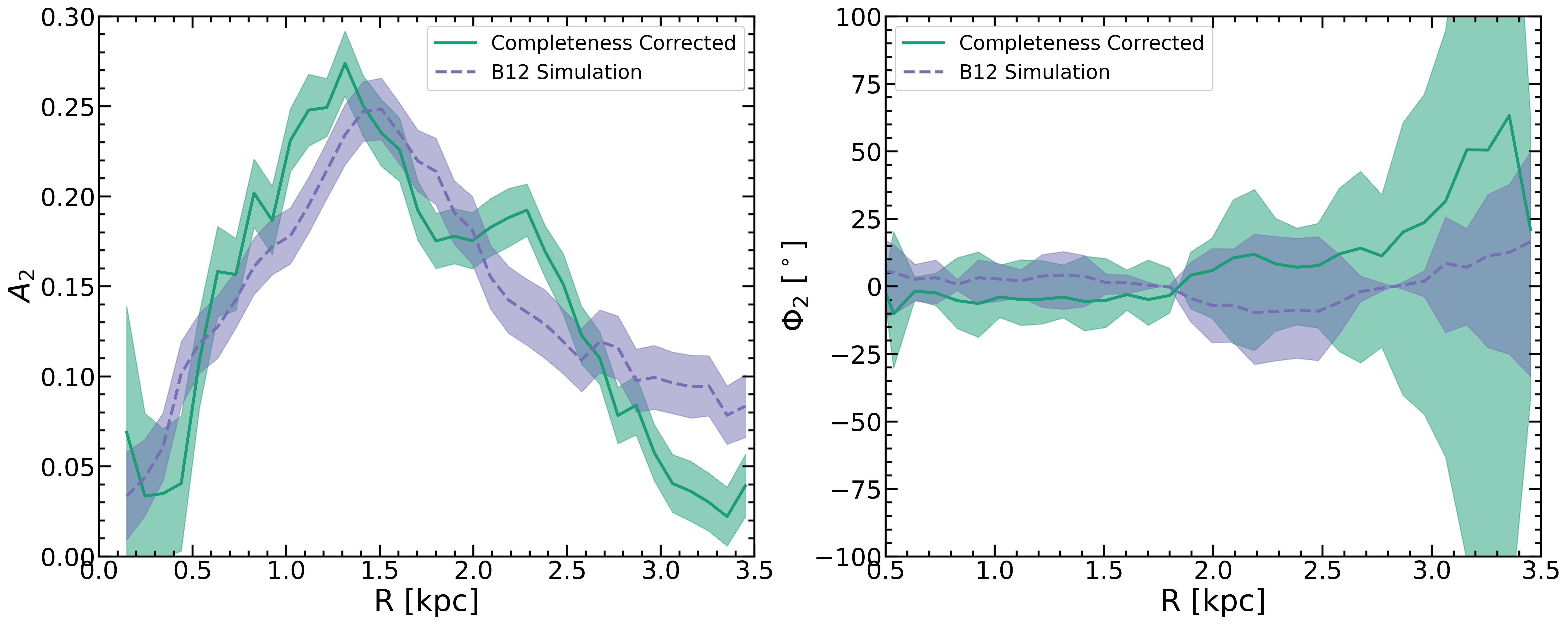}
    \caption{Similar to Figure \ref{fig:bar_fourier}, but showing the comparison between the $m = 2$ Fourier amplitude and phase profiles of the completeness corrected Gaia observations (green, solid line) and the B12 simulation in which the SMC has recently collided with the LMC (blue, dashed line). The agreement between the simulation and observations affirms the validity of such a model.}
    \label{fig:bar_fourier_with_b12}
\end{figure*}

We perform a Fourier decomposition on the stellar particles in the LMC B12 Model-2 present day snapshot, and determine the simulated stellar bar's parameters. We use the same procedure for the simulation as was done for observations, the only difference being that the simulated LMC disk does not require a completeness correction. The radial profile of the bi-symmetric Fourier amplitude ($A_2$) for the simulations and the completeness corrected observations is shown in Figure \ref{fig:bar_fourier_with_b12}. The simulations and data match qualitatively (the width of the shaded region is the $3-\sigma$ bootstrapped error) and quantitatively (Table \ref{tab:bar_props}). 

We calculate the bar offset in the B12 simulation using the same procedure as was done for observations. We find the outer disk radius for the simulation to be 7.41 kpc. We find a bar offset in the B12 simulation as well (Figure \ref{fig:bar_ellip}, right panel), which is a hallmark of close galaxy interactions \citep{Besla2016, Pardy2016}. However, the offset in the B12 simulation is larger compared to the observations, which likely means that the timing of the SMC's close passage in the simulation with respect to the present day snapshot is not correct. This is in agreement with the results of C22, who compared the level of disk heating of the observed LMC with the B12 simulations. Thus, accurate timing of the SMC's interactions with the LMC is a limitation of existing simulations of the Clouds, which we plan to resolve in the future with updated simulations.

We report the bar parameters in the B12 simulation in Table \ref{tab:bar_props}. We find general agreement between the bar parameters in the B12 simulation and the completeness corrected observations. This suggests that the B12 simulation is a reasonable representation of the LMC bar, which is consistent the hypothesis of a recent direct collision between the SMC and the LMC.

Note that the  simulations of the LMC-SMC-MW interaction history used to compare against our observations (B12, Model-2) do not reproduce some key observables of the LMC-SMC system, like the SMC’s present day Galactocentric position and velocity and the LMC’s disk inclination. However, our goal is primarily to show that the observed LMC bar structure is consistent with a scenario where the SMC has undergone a recent collision with the LMC. The simulations are well-suited to study this scenario in general. New simulations that are better matched to observations are needed for the purpose of an in-depth comparative study, which is work in progress. Future work also involves comparison of the observed bar properties with a library of simulated LMC-like disks under a variety of interaction scenarios for LMC and SMC like galaxies \citep[e.g.][KRATOS simulations]{KRATOS2024}.

\subsection{Placing the LMC in Context with Other Barred Galaxies} \label{sec:context}

As described in section \ref{sec:intro}, the LMC's bar is a very strange bar compared to bars in other galaxies. In this work, having measured several bar parameters precisely, we are in a position to directly compare the LMC bar's properties to bars in other galaxies. To the best of our knowledge, such a comparison has not been done before for the LMC.

\cite{Cuomo2020, Cuomo2022, Cuomo2024} measured the bar properties of a sample of 104 galaxies in the nearby universe primarily comprising of observations from the SDSS-MaNGA Integral Field Unit (IFU) survey \citep{Bundy2015} and the CALIFA IFU survey \citep{Sanchez2012}. Their sample (hereafter referred to as the Cuomo+ sample) comprises of a wide range of luminosities and morphological types. \cite{Cuomo2020} study scaling relations of the bar radius ($R_{bar}$) with the bar strength ($S_{bar}$) and with the galaxy R-band absolute magnitude ($M_r$). The R-band absolute magnitude can be used as a representative of the galaxy mass.

Longer bars generally tend to be stronger \citep{Erwin2005, Guo2019}. This trend is a consequence of the evolution of the bar. With time, bars exchange angular momentum with other components of the galaxy like the dark matter halo, and grow in radius and strength as they trap more stars in this process  \citep{Debattista1998, Athanassoula2013}. \cite{Sheth2008} attribute the trend of the bar radius with luminosity/mass to a downsizing process, where massive galaxies tend to form bars earlier in time, and the bars are longer at present day as a result of secular evolution through the exchange of angular momentum. However, this hypothesis has been partially debated in the context of observational selection effects \citep{Erwin2018, Erwin2019}.

\begin{figure*}
    \includegraphics[width = 0.48\textwidth]{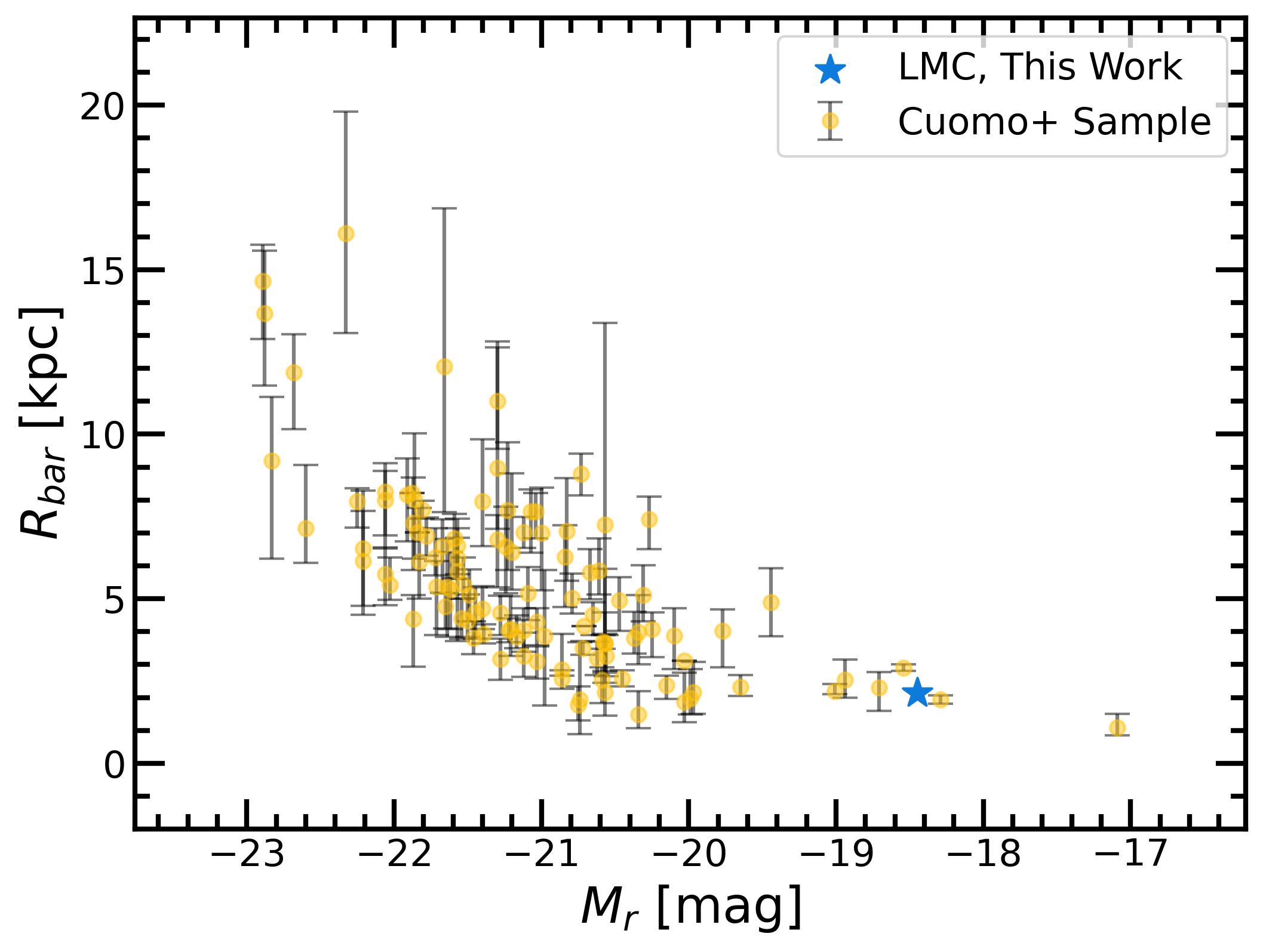}
    \includegraphics[width = 0.48\textwidth]{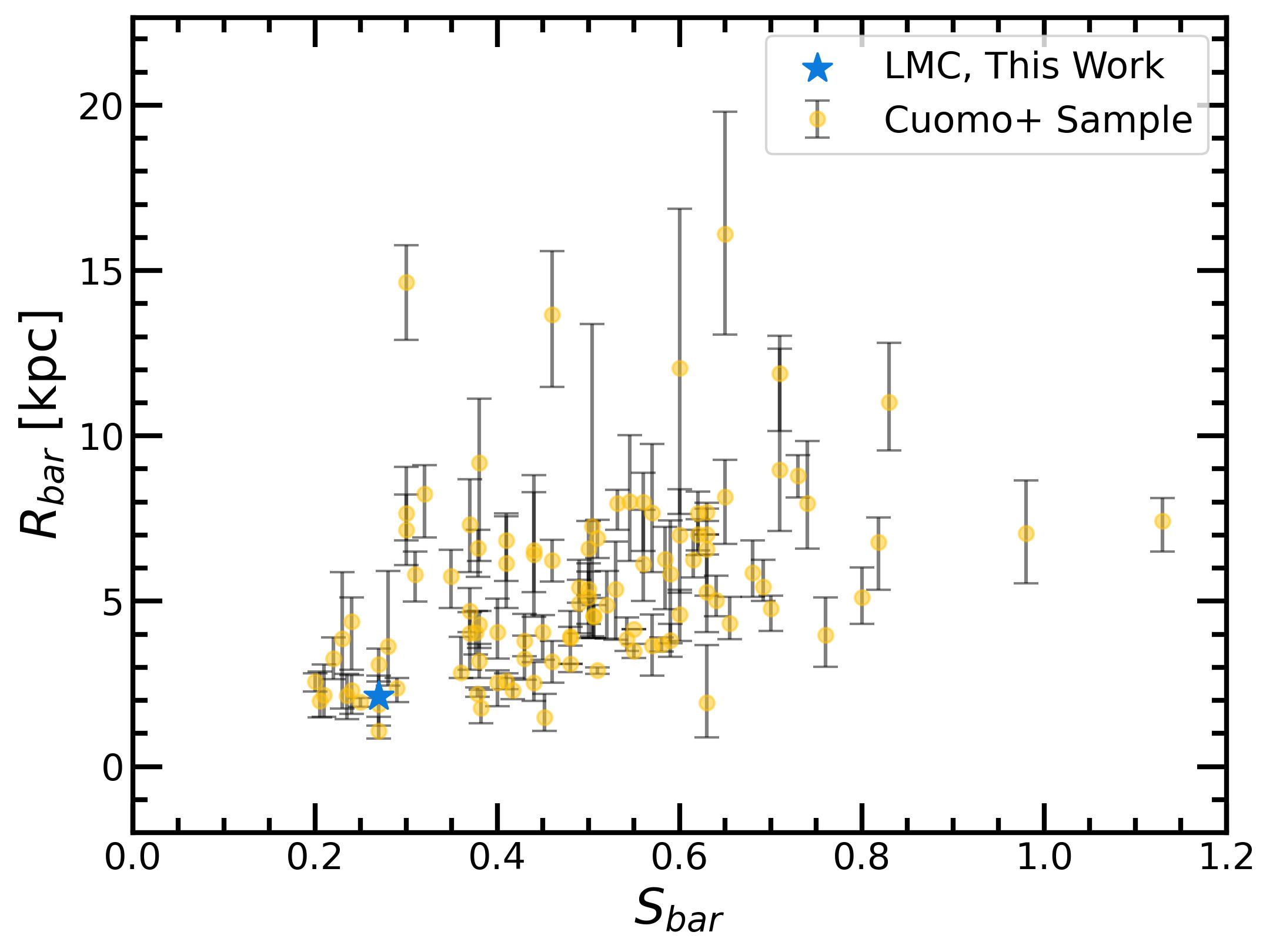}
    \caption{{\em Left panel}: scaling relation of the bar radius ($R_{bar}$) with the galaxy R-band absolute magnitude ($M_r$). Yellow points with errorbars are a diverse sample of galaxies from \cite{Cuomo2020, Cuomo2022, Cuomo2024}. The blue star is our measurements of the LMC. {\em Right panel}: scaling relation of the bar radius with the bar strength ($S_{bar}$). The LMC lies within the scaling relations of both Panel (a) and Panel (b), indicating that the LMC is not a strange galaxy from the viewpoint of the bar-galaxy connection.
    \label{fig:scaling}}
\end{figure*}

We place our measurements of the LMC in these scaling relations in Figure \ref{fig:scaling}. We compute the R-band absolute magnitude of the LMC from the R-band apparent magnitude \citep{deV1960} and the assumed distance of $49.9$ kpc. We find that the LMC resides within both the $R_{bar}$ - $M_r$ and $R_{bar}$ - $S_{bar}$ relations.

Thus, from the viewpoint of the bar-galaxy connection, the LMC is similar to other galaxies of a similar mass. The LMC also has a bar axis ratio within the observed range for star-forming galaxies \citep{Martin1995}. Yet, its bar shows strange properties like an offset, tilt and absence in the ISM. In a forthcoming paper (Rathore et al. 2024(b) in prep), we will show that these strange properties can be accounted for by the SMC's interactions with the LMC.

\subsection{Other Applications of the Completeness Correction Framework}

The completeness correction framework that we have developed in this work can be applied to other systems where measurements rely on star counts and crowding is a significant issue. For example, the SMC's central region \citep[e.g.][]{Bell2020}, Milky Way globular clusters \citep[e.g.][]{CantatGaudin2023, Imants2024} and the Milky Way's central regions \citep[e.g.][]{Dixon2023}. Such a framework is particularly necessary when other methods, such as artificial star tests, are computationally expensive or impractical for large datasets.
\\
\\
In the local universe several examples of LMC-like galaxies have been found. These galaxies are characterized by off-centered bars and one-armed spirals, and have a mass similar to the LMC \citep{deVFreeman72, Liu2011, Mao2024}. We posit that these LMC-like galaxies have had a similar evolutionary history as the LMC. Meaning, interactions with an SMC analog that is either in the process of merging, or has already merged, are responsible for the strange bar properties. Indeed, LMC-SMC pairs are not uncommon from a cosmological point of view \citep[e.g.][]{Besla2018, Chamberlain2024}. We will explore this hypothesis in detail in forthcoming papers.

\section{Conclusions} \label{sec:conclusion}

The stellar bar of the LMC has several strange properties. The bar is offset from the disk center, tilted with respect to the disk plane, and has no signature in the ISM. Precise characterization of the LMC bar's geometry (bar radius, axis ratio, position angle) and strength are necessary for discerning the origin of its strange properties and placing the LMC in context with other barred galaxies. 

We utilize observations of the LMC's red clump stars in Gaia DR3. Red clump stars constitute an old stellar population, thus they are an ideal tracer for the LMC's bar. They occupy a well defined region in the Hertzsprung Russel diagram, which makes them easy to select in observational datasets. Moreover, they have a narrow distribution of mass, which facilitates comparison with numerical simulations having star particles of equal masses.  

A major challenge to measuring the LMC's bar properties is crowding in the LMC's central regions. We propose a novel solution to correct for crowding induced incompleteness in stellar photometry by using the Gaia BP-RP color excess. Using the color excess, we derive a completeness map for the LMC's disk and find that completeness in the LMC's central region is underestimated by at least an order of magnitude in Gaia.

We use the method of Fourier decomposition combined with iso-ellipse fitting to precisely measure the LMC's bar. We find that incompleteness significantly increases the uncertainty in the inferred bar properties, which can potentially be the source of discrepancy between measurements in literature. 

Through our completeness correction, we find the LMC's bar radius is $R_{bar} = 2.13^{+0.03}_{-0.04}$ kpc and the bar position angle is $121.26^{\circ} \pm 0.21^{\circ}$. The LMC's bar has an axis ratio of $0.54 \pm 0.03$, meaning it is thicker than previous estimates. The LMC bar center is also offset with respect to the disk center by $0.76 \pm 0.01$ kpc; this is an important parameter for comparison to dynamical models of an LMC-SMC collision.

For the first time, we find the LMC bar's strength using the stellar density field, as quantified by the Fourier bi-symmetric amplitude, to be $S_{bar} = 0.27$. This value is significantly above the threshold ($S_{bar} = 0.20$) used to define barred galaxies, meaning the LMC is a traditional barred galaxy.  

Through this work, we have presented a reliable geometric framework that can accurately determine multiple bar parameters, without relying on assumptions of the mass profile of the bar and disk. We find a general agreement between the bar parameters measured from our framework and literature measurements that are based on similar methods \citep[e.g.][]{Arranz2024}.

We compare the observed bar properties with the \cite{Besla2012} Model-2 hydrodynamic model where the SMC has undergone a recent direct collision with the LMC. We find consistency between observations and the model, supporting the theory that the LMC was originally a symmetric, barred spiral galaxy that was then perturbed by the SMC through a collision. We further place the LMC in context of other barred galaxies in the local universe for the first time by studying scaling relations between bar and galaxy properties. We find that the LMC is similar to other galaxies from the viewpoint of bar-galaxy connection. We argue that the strange properties of the LMC's bar are primarily a result of the SMC's interactions. We highlight other systems where our completeness correction framework as well as the Fourier decomposition technique can be utilized to precisely measure morphology in the Local Group.

\begin{acknowledgements}
Himansh Rathore would like to acknowledge interesting discussions with Mathieu Renzo, Peter Behroozi, Leandro Beraldo-e-Silva, David Sand, Dennis Zaritsky, {\'O}. Jim{\'e}nez-Arranz, Ewa L. {{\L}okas}, Anthony Brown, Alfred Castro-Ginard, Tristan Cantat-Gaudin,  Lia Athanassoula, Roeland van der Marel and Kathryne J. Daniel regarding the LMC's bar as well as the completeness calibration framework. We would like to thank the anonymous referee for providing insightful comments that improved the quality as well as clarity of the paper. We would like to thank the journal data editor for suggestions which helped to add and improve software citations in the manuscript. This work utilized the Puma and ElGato High Performance Computing clusters at the University of Arizona.

Himansh Rathore and Gurtina Besla are supported by NSF CAREER AST 1941096, NASA ATP 80NSSC24K1225 and NASA FINESST 23-ASTRO23-0004.

This research uses services or data provided by the Astro Data Lab, which is part of the Community Science and Data Center (CSDC) Program of NSF NOIRLab. NOIRLab is operated by the Association of Universities for Research in Astronomy (AURA), Inc. under a cooperative agreement with the U.S. National Science Foundation.

This work has made use of data from the European Space Agency (ESA) mission
{\it Gaia} (\url{https://www.cosmos.esa.int/gaia}), processed by the {\it Gaia}
Data Processing and Analysis Consortium (DPAC,
\url{https://www.cosmos.esa.int/web/gaia/dpac/consortium}). Funding for the DPAC
has been provided by national institutions, in particular the institutions
participating in the {\it Gaia} Multilateral Agreement.

We respectfully acknowledge that the University of Arizona is on the land and territories of Indigenous peoples. Today, Arizona is home to 22 federally recognized tribes, with Tucson being home to the O’odham and the Yaqui. Committed to diversity and inclusion, the University strives to build sustainable relationships with sovereign Native Nations and Indigenous communities through education offerings, partnerships, and community service.

\facilities{Gaia, Astro Data Lab, UArizona HPC}
\software{This work made use of python, and its packages like numpy \citep{vanderWalt2011, Harris2020}, scipy \citep{Virtanen2020}, matplotlib \citep{Hunter2007} and astropy \citep{Astropy2013, Astropy2018, Astropy2022}. This work made use of the open source database management software topcat \citep{Taylor2005}.}
\end{acknowledgements}

\appendix

\section{Gaia Data Selection}
\begin{figure}
    \centering
    \includegraphics[width = 0.4\textwidth]{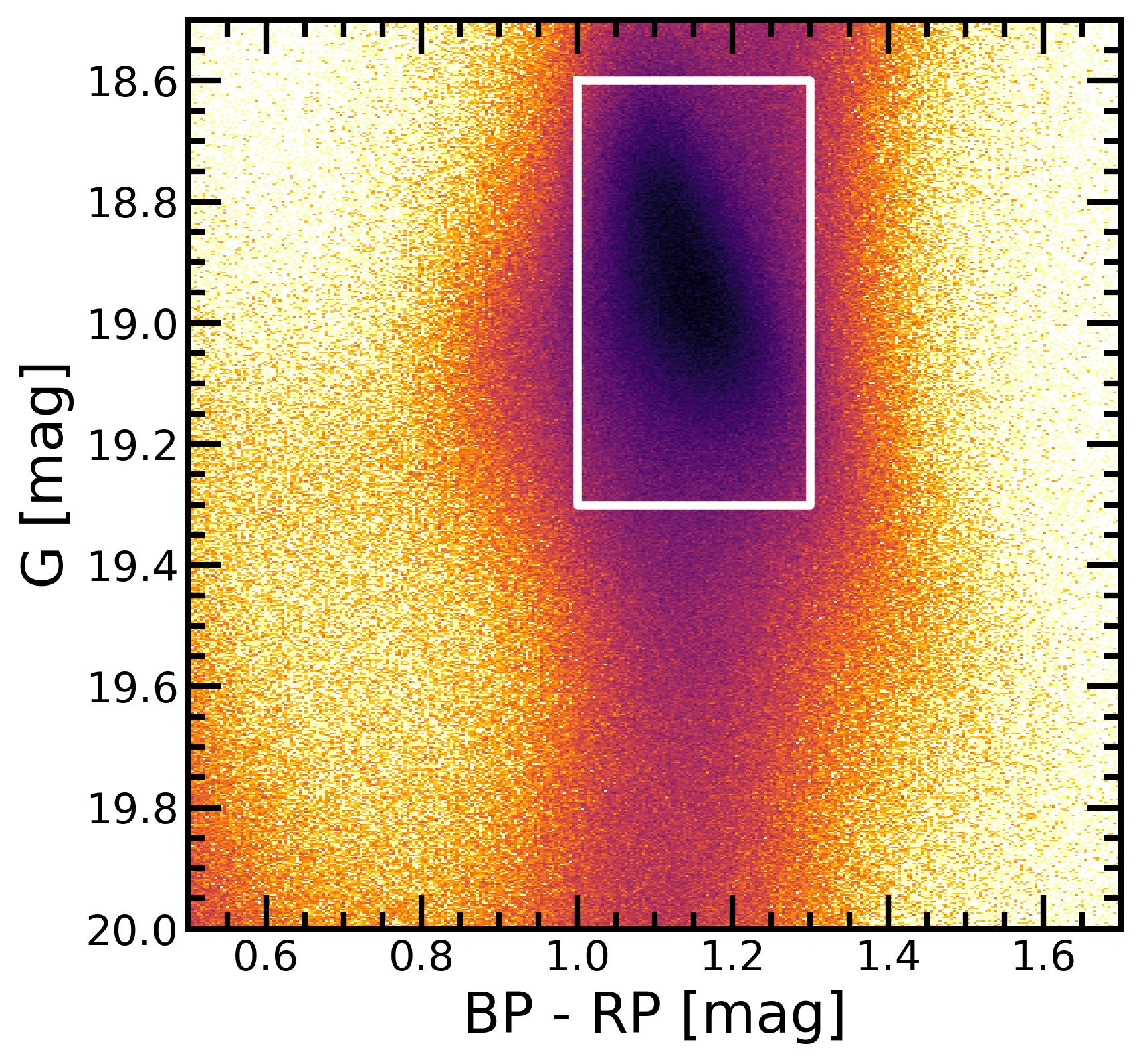}
    \caption{The color-magnitude diagram of a sample of LMC's stars identified in Gaia DR3. The white box shows our selection area for red clump stars.}
    \label{fig:cmd}
\end{figure}

Here, we outline our selection procedure for the LMC's red clump stars in Gaia DR3 (main text: section \ref{sec:gaiadata}). We use the following criteria:
\begin{enumerate}
    \item $50^\circ \leq \alpha \leq 110^\circ$ and $-80^\circ \leq \delta \leq -55^\circ$
    \item $\varpi < 0.1$ and $\frac{\varpi}{\sigma_\varpi} < 5 $
    \item $\left(\mu_{\alpha\ast} - 1.8593\right)^2 + \left(\mu_{\delta} - 0.3747\right)^2 < 1.5^2$
    \item astrometric\_excess\_noise $< 0.2$
    \item phot\_bp\_mean\_mag - phot\_rp\_mean\_mag $> 0.5$
    \item phot\_g\_mean\_mag $> 18.4$
    \item $\rho < 10^\circ$   
\end{enumerate}
\noindent where $\varpi$ is the parallax in mas; $\mu_{\alpha\ast}$ and $\mu_\delta$ are the proper motions in Right Ascension (RA; $\alpha$, $\alpha^\ast = \alpha \cos\delta$) and declination (DEC; $\delta$) in mas yr$^{-1}$ respectively; $\rho$ is the galactic radius from the LMC kinematic center $(\alpha_0, \delta_0) = (80.443^\circ, -69.272^\circ)$ as determined in C22. We use the following query in the Gaia query interface:
\\
\\
\enquote{SELECT source\_id, ra, dec, phot\_g\_mean\_flux, phot\_g\_mean\_flux\_error, phot\_g\_mean\_flux\_over\_error
FROM \\ gaia\_dr3.gaia\_source WHERE ra $>= 50$ AND ra $<= 110$ AND dec $>= -80$ AND dec $<= -55$ AND parallax $< 0.1$ AND parallax\_over\_error $< 5$ AND 't' $=$ Q3C\_RADIAL\_QUERY(pmra, pmdec, 1.8593, 0.3747, 1.5) AND astrometric\_excess\_noise $< 0.2$ AND phot\_bp\_mean\_mag $-$ phot\_rp\_mean\_mag $> 0.5$ AND phot\_g\_mean\_mag $> 18.4$ AND 't' $=$ Q3C\_RADIAL\_QUERY(ra, dec, 80.443, -69.272, 10)}
\\
\\
We show the color-magnitude diagram of the sample obtained using the above query in Figure \ref{fig:cmd}. We select red clump stars from the color-magnitude diagram using the selection criteria of C22 (denoted by the white box in Figure \ref{fig:cmd}):

$$(G, BP-RP) = (1.0, 18.6), (1.3, 18.6), (1.3, 19.3), (1.0, 19.3)$$

\section{Coordinate Systems}
Here we outline the conversions between various coordinate systems used in this work to describe the LMC's disk (main text: section \ref{sec:coords}). 

The LMC tangent plane coordinates can be obtained from the equitorial RA ($\alpha$) - DEC ($\delta$) coordinates by the following transformation:

\begin{equation}
    \xi = \frac{\cos(\delta)\sin(\alpha - \alpha_0)}{\sin(\delta_0)\sin(\delta) + \cos(\delta_0)\cos(\delta)\cos(\alpha - \alpha_0)}
\end{equation}

\begin{equation}
    \eta = \frac{\cos(\delta_0)\sin(\delta) - \sin(\delta_0)\cos(\delta)\cos(\alpha - \alpha_0)}{\sin(\delta_0)\sin(\delta) + \cos(\delta_0)\cos(\delta)\cos(\alpha-\alpha_0)}
\end{equation}

We set the center $(\alpha_0, \delta_0) = (80.443^\circ, -69.272^\circ)$ as the LMC kinematic center obtained in C22. 

The LMC in-plane cartesian coordinates can be obtained by the following transformation:

\begin{equation}
x = D\sin(\rho)cos(\phi - \theta)
\end{equation}
\begin{equation}
y = D\left[\sin(\rho)\sin(\phi - \theta)\cos(i) + \cos(\rho)\sin(i)\right] - D_0\sin(i)
\end{equation}
\begin{equation}
z = D\left[\sin(\rho)\sin(\phi - \theta)\sin(i) - \cos(\rho)\cos(i)\right] + D_0\cos(i)
\end{equation}

\noindent where,
\begin{equation}
    D = \frac{D_0\cos(i)}{\cos(i)\cos(\rho) - \sin(i)\sin(\rho)\sin(\phi - \theta)}
\end{equation}

\noindent $D_0 = 49.9$ kpc is the distance to the LMC center \citep{deGrijs2014}. $(i, \theta) = \left( 25.86^\circ, 149.23^\circ \right)$ are the inclination and line-of-nodes position angle respectively \citep{Choi2018a}. 
$(\rho, \phi)$ are the referred to as the LMC angular coordinates, which are calculated as follows:

\begin{equation}
\cos(\rho) = \sin(\delta)\sin(\delta_0) + \cos(\delta)\cos(\delta_0)\cos(\alpha - \alpha_0)
\end{equation}
\begin{equation}
\tan(\phi) = \frac{\cos(\delta)\sin(\delta_0)\cos(\alpha - \alpha_0) - \sin(\delta)\cos(\delta_0)}{\cos(\delta)\sin(\alpha - \alpha_0)}
\end{equation}

\bibliography{references}{}
\bibliographystyle{aasjournal}

\end{document}